\documentclass[floatfix,showpacs,amsmath,amssymb,preprint,prd,12pt]{revtex4}
\usepackage{graphicx}
\begin{document}
\title{Next-to-leading order QCD effects
in associated charged Higgs and W boson production in the MSSM at
the CERN Large Hadron Collider}
\author{Jun Gao}
\email{gaojun49@pku.edu.cn}
\author{Chong Sheng Li}
\email{csli@pku.edu.cn}
\author{Zhao Li}
\email{zhli.phy@pku.edu.cn}
\affiliation{Department of Physics, Peking University, Beijing 100871, China}

\begin{abstract}
We present the calculations of the next-to-leading order (NLO) QCD
corrections to the inclusive total cross sections for the associated
production of the $W^{\pm}H^{\mp}$ through $b\bar{b}$ annihilation
in the Minimal Supersymmetric Standard Model at the CERN Large
Hadron Collider. The NLO QCD corrections can either enhance or
reduce the total cross sections, but they generally efficiently
reduce the dependence of the total cross sections on the
renormalization/factorization scale. The magnitude of the NLO QCD
corrections is about 10\% in most of the parameter space and can
reach 15\% in some parameter regions. We also show the Monte Carlo
simulation results for the $2j+\tau_{jet}+\not{\! p}_{T}$ signature
from the $W^{\pm}$ and the $H^{\mp}$ decays including the NLO QCD
effects, and find an observable signal at a $5\sigma$ level in some
parameter region of the minimal supergravity model.
\end{abstract}
\maketitle

\section{INTRODUCTION}

The Higgs mechanism~\cite{Higgs:1964ia} plays a key role for the
understanding of the spontaneous electroweak symmetry breaking in
both the Stand Model (SM) and the Minimal Supersymmetric Stand Model
(MSSM)~\cite{Nilles:1983ge}. Searching for Higgs bosons is one of
the most important missions for the upcoming CERN Large Hadron
Collider (LHC). The MSSM contains five physical Higgs bosons: two
neutral CP-even bosons $h^{0}$ and $H^{0}$, one neutral CP-odd boson
$A^{0}$, and the charged $H^{\pm}$ boson pair. The $h^{0}$ is the
lightest and SM-like Higgs boson, while the others are non-SM-like
ones whose discovery will give the direct evidence of new physics
beyond the SM, especially charged Higgs boson.

At hadron colliders, the charged Higgs bosons $H^{\pm}$ could appear
as the decay product of primarily produced top quarks if the mass of
$H^{\pm}$ is smaller than $m_{t}-m_{b}$. For heavier $H^{\pm}$,
single charged Higgs boson production associated with heavy quark,
such as $gb \to H^{-}t$~\cite{Bawa:1989pc}, $qb \to
q^{'}bH^{-}$~\cite{Moretti:1996ra}, and $q\bar{q},gg \to
tbH^{\pm}$~\cite{Hesselbach:2007jj}, are the main channels for
single charged Higgs boson production. The channels for pair
production are $q\bar{q}$ annihilation and the loop-induced $gg$
fusion process~\cite{Eichten:1984eu}. These processes have large
production rates, but also suffer from large QCD backgrounds,
especially when the $H^{\pm}$ mass is larger than $m_{t}+m_{b}$.
Another attractive channel is single charged Higgs boson production
associated with W boson~\cite{Dicus:1989vf}. The dominant partonic
subprocesses at the LHC are $b\bar{b} \to W^{\mp}H^{\pm}$ at the
tree-level and $gg \to W^{\mp}H^{\pm}$ at the one-loop
level~\cite{Bendezu:2000tu}. For the $b\bar{b}$ annihilation
process, the supersymmetric electroweak (SUSY-EW), the
$\mathcal{O}(\alpha_{s})$ pure QCD and the supersymmetric QCD
(SUSY-QCD) corrections have been calculated in
Ref.~\cite{Yang:2000yt}~\cite{Hollik:2001hy}~\cite{Zhao:2005mu},
respectively. In this paper, we use the dimensional reduction
(DRED)~\cite{Bern:2002zk} scheme to regularize both the ultraviolet
(UV) and the infrared (IR) divergences while in
Ref.~\cite{Hollik:2001hy} the gluon was given a finite small mass to
regularize IR divergences. We will focus on the case of $\mu>0$
which is favored by the recent measurement of the anomalous magnetic moment of
the muon~\cite{Ellis:2001yu}, $\mu$ is the Higgs superfield mass term
in the superpotential, so the SUSY-QCD corrections
are relatively small and can be neglected as shown in
Ref.~\cite{Zhao:2005mu}. For simplicity, in our calculations, we
neglect the bottom quark mass except in the Yukawa couplings. Such
approximations are valid in all diagrams, in which the bottom quarks
appear as initial state partons, according to the simplified
Aivazis-Collins-Olness-Tung (ACOT) scheme~\cite{Aivazis:1993pi}.
Moreover, we only consider the process $b\bar{b} \to H^-W^+$ since
the cross section for the process $b\bar{b} \to H^+W^-$ is the
same if we choose all the relevant parameters to be real.

Recently, in Ref.~\cite{Eriksson:2006yt} the authors investigated
the viability of observing charged Higgs bosons produced in
association with W bosons at the LHC at LO level, using the leptonic
decay $H^{-} \to \tau^{-}\nu_{\tau}$ and hadronic W decay. In this
paper we also give the Monte Carlo simulation results of the above
signal, but in the minimal supergravity (mSUGRA)~\cite{Drees:1995hj}
scenario including the NLO QCD effects.

The arrangement of this paper is as follow. In Sec.~II, we show the
LO explicit expressions. In Sec.~III, we present the details of the
calculations for both the virtual and real QCD corrections. In
Sec.~IV, we give some analysis on the signal and background. Sec.~V
are the numerical results for total and differential cross sections
and the Monte Carlo simulation results. Sec.~VI contains a brief
conclusion. The relevant coupling constants and the lengthy analytic
expressions are summarized in Appendix.


\section{LEADING ORDER CALCULATIONS}

\begin{figure}[!ht]
\includegraphics[width=0.90\textwidth]{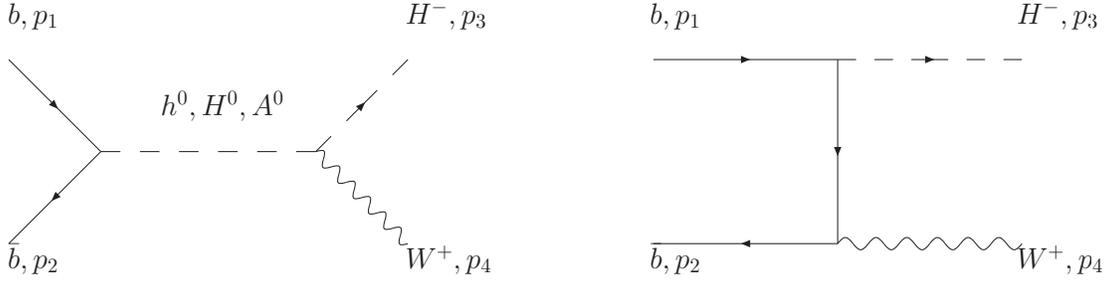}
\caption[]{Leading order Feynman diagrams for $b\bar{b} \to H^-W^+$.}
\label{f177}
\end{figure}

The tree-level Feynman diagrams for the subprocess $b(p_{1})\bar{b}(p_{2}) \to
H^{-}(p_{3})W^{+}(p_{4})$ are shown in Fig.\ref{f177},
and its LO amplitude in $n=4-2\epsilon$ dimension is
\begin{equation}
M^{B}=M^{(s)}_{h^{0}}+M^{(s)}_{H^{0}}+M^{(s)}_{A^{0}}+M^{(t)}_{0},
\end{equation} with
\begin{align}
  M^{(s)}_{h_{0}}&=-\frac{e^{2}m_{b}s_{\alpha}c_{\beta-\alpha}}{2c_{\beta}M_{W}s^{2}_{W}(s-m^{2}_{h^{0}})}(M_{5}+M_{6}+M_{9}+M_{10}),\nonumber\\
  M^{(s)}_{H_{0}}&=-\frac{e^{2}m_{b}c_{\alpha}s_{\beta-\alpha}}{2c_{\beta}M_{W}s^{2}_{W}(s-m^{2}_{H^{0}})}(M_{5}+M_{6}+M_{9}+M_{10}),\nonumber\\
  M^{(s)}_{A_{0}}&=-\frac{e^{2}m_{b}t_{\beta}}{2M_{W}s^{2}_{W}(s-m^{2}_{A^{0}})}(M_{5}-M_{6}+M_{9}-M_{10}),\nonumber\\
  M^{(t)}_{0}&=-\frac{e^{2}}{2M_{W}s^{2}_{W}(t-m^{2}_{t})}[\frac{m^{2}_{t}}{t_{\beta}}M_{2}-m_{b}t_{\beta}(2M_{9}+M_{3})],
\end{align}
where $s_{\alpha} \equiv \sin\alpha$, $c_{\alpha} \equiv
\cos\alpha$, $s_{\beta-\alpha} \equiv \sin(\beta-\alpha)$,
$c_{\beta-\alpha} \equiv \cos(\beta-\alpha)$, $t_{\beta} \equiv
\tan\beta$. Mandelstam variables $s, t$, and $u$ are defined as
follows:
\begin{equation}
  s=(p_{1}+p_{2})^{2}, \quad t=(p_{1}-p_{3})^{2}, \quad
  u=(p_{2}-p_{3})^{2}.
\end{equation}
$M_{i}$'s are reduced standard matrix elements, which are defined by\\
$\displaystyle{\hspace{4cm}M_{1(2)}=\bar{v}(p_{2})
\not{\!\epsilon}(p_{4})P_{R(L)}u(p_{1}),}$\\
$\displaystyle{\hspace{4cm}M_{3(4)}=\bar{v}(p_{2})
\not{\!p}_{4}\not{\!\epsilon}(p_{4})P_{R(L)}u(p_{1}),}$\\
$\displaystyle{\hspace{4cm}M_{5(6)}=\bar{v}(p_{2})
P_{R(L)}u(p_{1})p_{1}\cdot\epsilon(p_{4}),}$\\
$\displaystyle{\hspace{4cm}M_{7(8)}=\bar{v}(p_{2})
\not{\!p}_{4}P_{R(L)}u(p_{1})p_{1}\cdot\epsilon(p_{4}),}$\\
$\displaystyle{\hspace{4cm}M_{9(10)}=\bar{v}(p_{2})
P_{R(L)}u(p_{1})p_{2}\cdot\epsilon(p_{4}),}$\\
and\\
\begin{equation}
M_{11(12)}=\bar{v}(p_{2})
\not{\!p}_{4}P_{R(L)}u(p_{1})p_{2}\cdot\epsilon(p_{4}),\hspace{1.5cm}
\label{mmm}
\end{equation}
with the projectors $P_{L,R}\equiv(1\mp\gamma_{5})/2$ .

The LO total cross section at the LHC is obtained by convoluting
the partonic cross section with the parton distribution
functions (PDFs) $G_{b,\bar{b}/p}$ in the proton:
\begin{eqnarray}
\sigma^B=\int dx_1dx_2
[G_{b/p}(x_1,\mu_f)G_{\bar{b}/p}(x_2,\mu_f)+ (x_1\leftrightarrow
x_2)]\hat{\sigma}^{B},\label{Born0}
\end{eqnarray}
where $\mu_f$ is the factorization scale and $\hat{\sigma}^{B}= \int
\frac{1}{2s}\overline{\sum} |M^B|^2 d\Gamma$ is the Born level cross
section for $b(p_1) \bar{b}(p_2) \rightarrow H^-(p_3)W^+(p_4)$, in
which the colors and spins of the outgoing particles have been
summed, and the colors and spins of the incoming ones have been
averaged over.

\section{NEXT-TO-LEADING ORDER CALCULATIONS}
The NLO QCD contributions to the associated production of $H^{-}$
and $W^{+}$ through $b\bar{b}$ annihilation process consist of the
virtual corrections, generated by loop diagrams of colored
particles, and the real corrections with the radiation of a real
gluon or a massless (anti)bottom quark. For both virtual and real
corrections, we use DRED scheme to regularize all the divergences.
\subsection{Virtual corrections}
The virtual corrections to $b\bar{b} \to H^{-}W^{+}$ arise from the
Feynman diagrams shown in Fig.\ref{f18}, which consist of vertex,
self-energy and box diagrams. We carried out the calculation in
't~Hooft-Feynman gauge and used the dimensional reduction in
$n=4-2\epsilon$ dimensions to regularize the ultraviolet, soft and
collinear divergences in the virtual loop corrections. In order to
remove the UV divergences, we use the modified minimal subtraction
$(\overline{\rm MS})$ scheme to renormalize the bottom quark mass
and wave function, while for the top quark mass and wave function we
use both the $\overline{\rm MS}$ scheme and the on-shell (OS) scheme
and compare them. Denoting $m_{b0}$, $m_{t0}$, $\psi_{b0}$ and
$\psi_{t0}$ as the bare quark masses and the bare wave functions,
respectively, the relevant renormalization constants $\delta m_{b}$,
$\delta m_{t}$, $\delta Z_{bL,R}$ and $\delta Z_{tL,R}$ are then
defined as
\begin{align}
  &m_{b0}=m_{b}+\delta m_{b},\nonumber\\
  &m_{t0}=m_{t}+\delta m_{t},\nonumber\\
  &\psi_{b0}=(1+\delta Z_{bL})^{1/2}\psi_{bL}+(1+\delta Z_{bR})^{1/2}\psi_{bR},\nonumber\\
  &\psi_{t0}=(1+\delta Z_{tL})^{1/2}\psi_{tL}+(1+\delta Z_{tR})^{1/2}\psi_{tR}.
\end{align}
with $\psi_{(b,t)L}=P_L\psi_{(b,t)}$
and $\psi_{(b,t)R}=P_R\psi_{(b,t)}$.
After calculating the self-energy diagrams in Fig.\ref{f18},
we obtain the explicit expressions for
all the renormalization constants as follows:
\begin{align}
  &\frac{\delta m_{b}^{\overline{\rm MS}}}{m_{b}}=\frac{\delta m_{t}^
  {\overline{\rm MS}}}{m_{t}}=-\frac{\alpha _{s}}
  {4\pi}3C_{F}\Delta,\nonumber\\
  &\delta Z_{L(R)b}^{\overline{\rm MS}}=\delta Z_{L(R)t}^{\overline
  {\rm MS}}=-\frac{\alpha
  _{s}}{4\pi}C_{F}\Delta,\nonumber\\
  &\frac{\delta m_{t}^{\rm OS}}{m_{t}}=-\frac{\alpha _{s}}
  {4\pi}C_{F}\left(\frac{4\pi\mu^2}{m^2}\right)^{\epsilon}\Gamma(1+\epsilon)
  \left(\frac{3}{\epsilon}+5\right),\nonumber\\
  &\delta Z_{L(R)t}^{\rm OS}=-\frac{\alpha
  _{s}}{2\pi}C_{F}\left[{\rm Re}(B_0+B_1)-2m_t^2{\rm Re}(B_0'-B_1')\right](m_t^2,0,m_t^2),
\end{align}
where $\displaystyle{\Delta \equiv \frac{1}{\epsilon}-\gamma_{E}+\ln
(4\pi)}$, $C_{F}=4/3$, and $B'=\partial B/\partial p^2$, $B_i$ are
the scalar two-point integrals~\cite{Denner:1991kt}.

The renormalized virtual amplitude can be written as
\begin{equation}
  M^{V}=M^{unren}+M^{con}.
\end{equation}
Here $M^{unren}$ contains the radiative corrections from the
one-loop vertex, self-energy and box diagrams, as shown in
Fig.\ref{f18}, and $M^{con}$ is the corresponding counterterm.
Moreover, $M^{unren}$ can be separated into two parts:
\begin{equation}
  M^{unren}=\sum_{\alpha=a}^{f}M^{\alpha}+M^{box},
\end{equation}
where $\alpha$ denotes the corresponding diagram indexes in
Fig.\ref{f18}. Using the standard matrix elements from Eq.(\ref{mmm})
they can be further expressed as
\begin{align}
  &M^{\alpha}=\sum_{l=1}^{12}f_{l}^{\alpha}M_{l},\nonumber\\
  &M^{box}=\sum_{l=1}^{12}f_{l}^{box}M_{l},\label{e5}
\end{align}
where $f_{l}^{\alpha}$ and $f_{l}^{box}$ are the form factors, which
are given explicitly in Appendix. The counterterm contribution
$M^{con}$ is separated into $M^{con(s)}$ and $M^{con(t)}$, i.e. the
counterterms for s and t channels, respectively, which are given by
\begin{eqnarray}
  &M^{con}\quad=&M^{con(s)}+M^{con(t)},\nonumber\\
  &M^{con(s)}=&(\frac{\delta m_{b}}{m_{b}}+\delta Z_{b})(M^{(s)}_{h^{0}}+M^{(s)}_{H^{0}}+M^{(s)}_{A^{0}}),
  \nonumber\\
  &M^{con(t)}=&M^{con(t)}_{1}+M^{con(t)}_{2},\nonumber\\
  &M^{con(t)}_{1}=&-\frac{e}{\sqrt {2}s_{w}(t-m^{2}_{t})^{2}}\left\{[m_{t}(t-m_{t}^{2})
  \delta Z_{t} -(m_{t}^{2}+t)\delta m_{t}]\right. \nonumber\\
  & & {} \left. aM_{2}-b[(t-m_{t}^{2})\delta Z_{t}-2m_{t}\delta m_{t}](M_{3}+2M_{9})\right\} ,\nonumber\\
  &M^{con(t)}_{2}=&M_{(0)}^{t}(\frac{\delta m_{b}}{m_{b}}+\delta Z_{b}+\delta
  Z_{t}),
\end{eqnarray}
with
\begin{equation}
  a=\frac{em_{t}}{\sqrt{2}M_{W}s_{W}t_{\beta}},\quad b=\frac{em_{b}t_{\beta}}{\sqrt{2}M_{W}s_{W}}.
\end{equation}
After adding all the terms above, the renormalized amplitude $M^{V}$
is UV finite, but still contains the IR divergences, and is given
by:
\begin{equation}
  M^{V}|_{IR}=\frac{\alpha _{s}}{2\pi}\frac{\Gamma (1-\epsilon)}{\Gamma (1-2\epsilon)}
  \left(\frac{4\pi\mu _{r}^{2}}{s}\right)^{\epsilon}\left(\frac{A_{2}^{V}}{\epsilon ^{2}}+\frac{A_{1}^{V}}{\epsilon}\right)M^{B},
\end{equation}
with
\begin{equation}
 A_{2}^{V}=-C_{F}, \quad A_{1}^{V}=-\frac{3}{2}C_{F}.
\end{equation}
Here the IR divergences include both the soft and the collinear
divergences. The soft divergences are canceled after adding the real
emission corrections, and the remaining collinear divergences can be
absorbed into the redefinition of PDF~\cite{Altarelli:1979ub}, which
will be discussed in the following subsections.
\subsection{Real gluon emission}
\begin{figure}[!ht]
\includegraphics[width=0.90\textwidth]{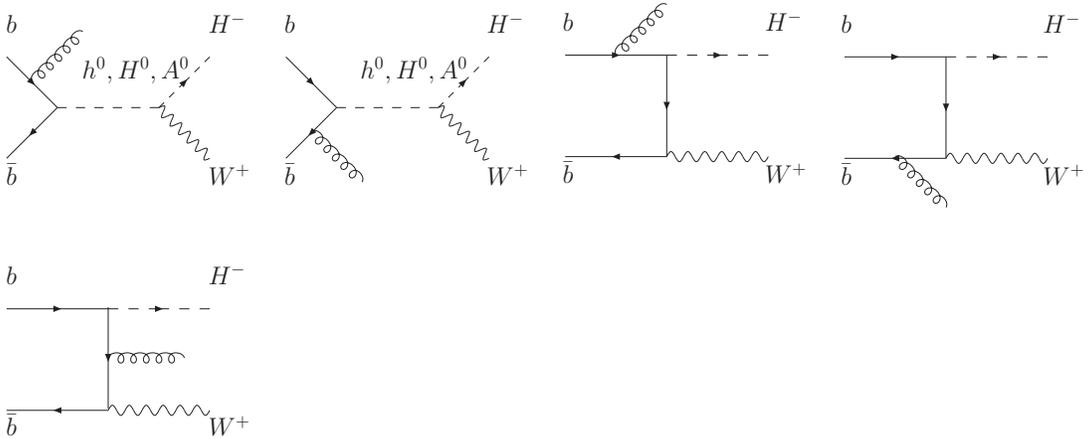}
\caption[]{Feynman diagrams for the real gluon emission contributions.}
\label{f19}
\end{figure}

The Feynman diagrams for the real gluon emission process $b(p_{1})\bar{b}(p_{2}) \to
H^{-}(p_{3})W^{+}(p_{4})+g(p_{5})$ are shown in Fig.\ref{f19}.

The phase space integration for the real gluon emission will produce
both soft and collinear infrared singularities, which can be
conveniently isolated by slicing the phase space into different
regions defined by suitable cutoff parameters. In this paper, we use
the two-cutoff phase space slicing method~\cite{Harris:2001sx},
which introduces two small cutoffs to decompose the three-body phase
space into three regions.

First, the phase space can be separated into two regions by an
arbitrary small cutoff $\delta _{s}$, according to whether the
energy $(E_{5})$ of the emitted gluon is soft, i.e. $E_{5}\le \delta
_{s} \sqrt {s}/2$, or hard, i.e. $E_{5}> \delta _{s}\sqrt {s}/2$.
Correspondingly, the partonic real cross section can be written as
\begin{equation}
  \hat{\sigma}^{R}=\hat{\sigma}^{S}+\hat{\sigma}^{H},
\end{equation}
where $\hat{\sigma}^{S}$ and $\hat{\sigma}^{H}$ are the
contributions from the soft and hard regions, respectively.
$\hat{\sigma}^{S}$ contains all the soft divergences. Second, in
order to isolate the remaining collinear divergences from
$\hat{\sigma}^{H}$, we should introduce another arbitrary small
cutoff, called collinear cutoff $\delta _{c}$, to further split the
hard gluon phase space into two regions, according to whether the
Mandelstam variables satisfy the collinear condition $-\delta
_{c}s<u_{1,2} \equiv (p_{1,2}-p_{5})^{2}<0$ or not. Thus, we have
\begin{equation}
  \hat{\sigma}^{H}=\hat{\sigma}^{HC}+\hat{\sigma}^{\overline{HC}},
\end{equation}
where the hard collinear part $\hat{\sigma}^{HC}$ contains the
collinear divergences, while the hard noncollinear part
$\hat{\sigma}^{\overline{HC}}$ is finite and can be numerically
computed using standard Monte-Carlo integration techniques and can
be written as
\begin{equation}
 d\hat{\sigma}^{\overline{HC}}=\frac{1}{2s}\overline{\sum}| M^{b\bar{b}} | ^{2}
 d \bar{\Gamma}_{3}.
\end{equation}
Here $d\bar{\Gamma}_{3}$ is the hard non-collinear region of the
three-body phase space.

In the next two subsections, we will discuss in detail the soft and hard collinear gluon
emission.
\subsubsection{Soft gluon emission}
In the soft limit, i.e. when the energy of the emitted gluon is
small, with $E_{5}\le \delta_{s} \sqrt{s}/2$, the matrix element
squared $\overline{\sum}| M^{R} | ^{2}$ for the process
$b(p_{1})\bar{b}(p_{2}) \to H^{-}(p_{3})W^{+}(p_{4})g(p_{5})$ can be
simply factorized into the Born matrix element squared times an
eikonal factor $\Phi _{eik}$:
\begin{equation}
\overline{\sum}| M^{R}(b\bar{b}\to H^{-}W^{+}+g)| ^{2} \stackrel{soft}{\to}
(4\pi \alpha _{s} \mu ^{2\epsilon}_{r})\overline{\sum} | M^{B} | ^{2}
\Phi _{eik},
\end{equation}
where the eikonal factor $\Phi _{eik}$ is given by
\begin{equation}
 \Phi _{eik}=C_{F}\frac{s}{(p_{1} \cdot p_{5})(p_{2} \cdot p_{5})}.
\end{equation}
Moreover, the phase space in the soft limit can also be factorized as
\begin{equation}
 d\Gamma _{3}(b\bar{b} \to H^{-}W^{+}+g)\stackrel{soft}{\to}d\Gamma _{2}(b\bar{b}\to
  H^{-}W^{+})dS,
\end{equation}
where $dS$ is the integration over the phase space of the soft gluon,
which is given by~\cite{Harris:2001sx}
\begin{equation}
 dS=\frac{1}{2(2\pi)^{3-2\epsilon}} \int_{0}^{\delta _{s} \sqrt{s}/2}
 dE_{5}E_{5}^{1-2\epsilon}d\Omega _{2-2\epsilon}.
\end{equation}
Hence, the parton level cross section in the soft region can be expressed as
\begin{equation}
 \hat{\sigma}^{S}=(4\pi \alpha_{s} \mu _{r}^{2\epsilon})\int d \Gamma_{2}
 \overline{\sum} | M^{B} | ^{2} \int dS \Phi_{eik}.\label{g12}
\end{equation}
Using the approach of Ref.~\cite{Harris:2001sx}, after analytically
integrating over the soft gluon phase space, Eq.(\ref{g12}) becomes
\begin{equation}
 \hat{\sigma}^{S}=\hat{\sigma}^{B}\left[\frac{\alpha_{s}}{2\pi}\frac{\Gamma(1-\epsilon)}
 {\Gamma(1-2\epsilon)}\left(\frac{4\pi\mu_{r}^{2}}{s}\right)^{\epsilon}\right]
 \left(\frac{A_{2}^{s}}{\epsilon ^{2}}+\frac{A_{1}^{s}}{\epsilon}+A_{0}^{s}\right),
\end{equation}
with
\begin{equation}
 A_{2}^{s}=2C_{F},\quad A_{1}^{s}=-4C_{F} \ln \delta_{s},\quad A_{0}^{s}=4C_{F} \ln^{2} \delta_{s}.
\end{equation}
\subsubsection{Hard collinear gluon emission}
In the hard collinear region, i.e. $E_{5}>\delta_{s}\sqrt{s}/2$ and $-\delta_{c}s<u_{1,2}<0$, the
emitted hard gluon is collinear to one of the incoming partons. As a consequence of the factorization
theorems~\cite{Collins:1985ue},
the squared matrix element for $b\bar{b} \to H^{-}W^{+}+g$ can be factorized into
the product of the Born squared matrix element and the Altarelli-Parisi splitting function for
$(b\bar{b}) \to b(\bar{b})g$~\cite{Altarelli:1977zs,
Ellis:1980wv},i.e.
\begin{equation}
 \overline{\sum} | M^{R}(b\bar{b} \to H^{-}W^{+}+g)| ^{2} \stackrel{collinear}{\to}
 (4\pi \alpha _{s} \mu_{r}^{2\epsilon}) \overline{\sum} | M^{B}| ^{2}
 \left(\frac{-2P_{bb}(z)}{zu_{1}}+\frac{-2P_{\bar{b}\bar{b}}(z)}{zu_{2}}\right),
\end{equation}
where $z$ denotes the fraction of incoming parton $b(\bar{b})$'s
momentum carried by parton $b(\bar{b})$ with the emitted gluon
taking a fraction $(1-z)$, and $P_{ij}(z)$ are the usual
Altarelli-Parisi splitting kernels~\cite{Altarelli:1977zs}.
Explicitly,
\begin{align}
 P_{bb}(z)=P_{\bar{b}\bar{b}}(z)=C_{F}\frac{1+z^{2}}{1-z}+C_{F}\frac{3}{2}\delta(1-z).
\end{align}
Moreover, the three-body phase space can
also be factorized in the collinear limit, and, for example,
in the limit $-\delta_{c}s<u_{1}<0$ it has the following form~\cite{Harris:2001sx}:
\begin{equation}
 d\Gamma_{3}(b\bar{b} \to H^{-}W^{+}+g) \stackrel{collinear}{\to} d\Gamma_{2}(b\bar{b} \to
 H^{-}W^{+};s^{'}=zs)\frac{(4\pi)^{\epsilon}}{16\pi^{2}\Gamma (1-\epsilon)}dzdu_{1}[(z-1)u_{1}]^{-\epsilon}.
\end{equation}
Here the two-body phase space should be evaluated at the squared
parton-parton energy $zs$. Thus, the three-body cross section in the
hard collinear region is given by~\cite{Harris:2001sx}
\begin{align}
 d\sigma^{HC}=&\hat{\sigma}^{B}\left[\frac{\alpha _{s}}{2\pi}\frac{\Gamma (1-\epsilon)}
 {\Gamma(1-2\epsilon)}\left(\frac{4\pi \mu_{r}^{2}}{s}\right)^{\epsilon}\right]
 \left(-\frac{1}{\epsilon}\right)\delta _{c}^{-\epsilon}\left[P_{bb}(z)G_{b/p}(x_{1}/z)
 G_{\bar{b}/p}(x_{2})\right. \nonumber\\
 &\left. +P_{\bar{b}\bar{b}}(z)G_{\bar{b}/p}(x_{1}/z)G_{b/p}(x_{2})+(x_{1}\leftrightarrow x_{2})
 \right]\frac{dz}{z}\left(\frac{1-z}{z}\right)^{-\epsilon}dx_{1}dx_{2},\label{e2}
\end{align}
where $G_{b(\bar{b})/p}(x)$ is the bare PDF.
\subsection{Massless (anti)quark emission}
\begin{figure}[!ht]
\includegraphics[width=0.90\textwidth]{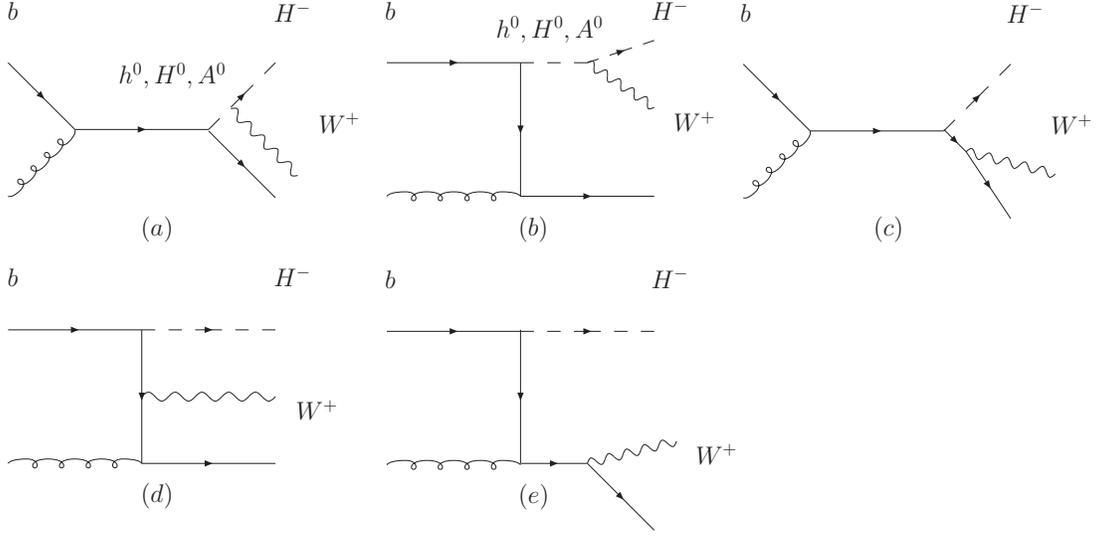}
\caption[]{Feynman diagrams for the emission of a massless bottom quark
contribution.}
\label{f20}
\end{figure}

In addition to the real gluon emission, a second set of real emission corrections to the inclusive
production rate of $pp \to H^{-}W^{+}$ at the NLO involves the processes with an additional
massless (anti)quark in the final states:
\begin{equation}
 gb \to bH^{-}W^{+},\quad g\bar{b} \to \bar{b} H^{-}W^{+}.\nonumber
\end{equation}
The relevant Feynman diagrams for massless (anti)quark emission (the
diagrams for the antiquark emission are similar and omitted here)
are shown in Fig.\ref{f20}.

Since the contributions from the real massless (anti)quark emission
contain the initial state collinear singularities, we also need to
use the two cutoff phase space slicing method~\cite{Harris:2001sx}
to isolate those collinear divergences. Because there is no soft
divergence in the splitting of $g \to b\bar{b}$, we only need to
separate the phase space into two regions: the collinear region and
the hard noncollinear region. Thus, according to the approach shown
in Ref.~\cite{Harris:2001sx}, the cross section for the processes
with an additional massless (anti)quark in the final states can be
expressed as
\begin{align}
 d\sigma ^{add}&=\sum _{(\alpha =g, \beta =b,\bar{b})} \hat{\sigma}^{\overline{C}}(\alpha \beta \to
 H^{-}W^{+}+X)[G_{\alpha /p}(x_{1})G_{\beta /p}(x_{2})+(x_{1}\leftrightarrow x_{2})]dx_{1}dx_{2}\nonumber\\
 &+\hat{\sigma}^{B}\left[\frac{\alpha_{s}}{2\pi}\frac{\Gamma (1-\epsilon)}{\Gamma (1-2\epsilon)}
 \left(\frac{4\pi \mu_{r}^{2}}{s}\right)^{\epsilon}\right](-\frac{1}{\epsilon})\delta_{c}^{-\epsilon}
 \left[P_{bg}(z)G_{g/p}(x_{1}/z)G_{\bar{b}/p}(x_{2})\right.\nonumber\\
 &\left.+G_{b/p}(x_{1})P_{\bar{b}g}(z)G_{g/p}(x_{2}/z)+(x_{1} \leftrightarrow x_{2})\right]
 \frac{dz}{z}\left(\frac{1-z}{z}\right)^{-\epsilon}dx_{1}dx_{2},\label{e1}
\end{align}
where
\begin{eqnarray}
&& P_{bg}(z)=P_{\bar{b}g}(z)=\frac{1}{2}[z^2 +(1-z)^2].
\end{eqnarray}
The first term in Eq.(\ref{e1}) represents the noncollinear cross
sections for the two processes, which can be written in the
form:
\begin{equation}
 d\hat{\sigma}^{\overline{C}}=\frac{1}{2s}\overline{\sum}| M^{\alpha \beta} | ^{2}
 d \overline{\Gamma}_{3},
\end{equation}
where $\alpha$ and $\beta$ denote the incoming partons in the
partonic processes, and $d\overline{\Gamma}_{3}$ is the three-body
phase space in the noncollinear region. The second term in
Eq.(\ref{e1}) represents the collinear singular cross sections.

Moreover, the top momentum in Fig.\ref{f20}(c) and (e) (as well as
in the corresponding $\bar{b}$ emission Feynman diagrams) can
approach the top mass shell, which will lead to a singularity
arising from the top propagator. Following the analysis shown in
Ref.~\cite{Beenakker:1996ch}, this problem can easily be solved by
introducing the non-zero top width $\Gamma _{t}$ and regularizing in
this way the higher-order amplitudes. However, these on-shell top
contributions are already accounted for by the LO level $tH^-$ and
$\bar{t}W^+$ productions with a subsequent decay, and thus should
not be considered as a genuine high-order correction to $H^-W^+$
associated production. Therefore, to avoid double counting, these
pole contributions will be subtracted in our numerical calculations
below in the same way as shown in Appendix B of
Ref.~\cite{Beenakker:1996ch}.

\subsection{Mass factorization}
As mentioned above, after adding the renormalized virtual
corrections and the real corrections, the partonic cross sections
still contain the collinear divergences, which can be absorbed into
the  redefinition of the PDF at NLO, in general called mass
factorization~\cite{Altarelli:1979ub}. This procedure in practice
means that first we convolute the partonic cross section with the
bare PDF $G_{\alpha/p}(x)$, and then rewrite $G_{\alpha/p}(x)$ in
terms of the renormalized PDF $G_{\alpha/p}(x,\mu_f)$. In the
$\overline{\rm MS}$ scheme and DRED scheme, the scale dependent PDF
$G_{\alpha/p}(x,\mu_f)$ is given by~\cite{Harris:2001sx}
\begin{align}
G_{\alpha/p}(x,\mu_f)=& G_{\alpha/p}(x)+
\sum_{\beta}(-\frac{1}{\epsilon})\bigg [\frac{\alpha_s}{2\pi}
\frac{\Gamma(1 -\epsilon)}{\Gamma(1 -2\epsilon)} \bigg(\frac{4\pi
\mu_r^2}{\mu_f^2}\bigg)^\epsilon\bigg]  \int_x^1 \frac{dz}{z}
P_{\alpha\beta}^+(z) G_{\beta/p}(x/z) \nonumber\\
&-\frac{\alpha _s}{2\pi} \sum_{\beta} \int_{x}^{1} \frac{dy}{y} P'_{\alpha\beta}(x/y)
G_{\alpha /p}(x),
\end{align}
where $P^+_{\alpha\beta}$ are the regulated splitting functions and
$P'_{ij}(z)$ are the usual Altarelli-Parisi splitting
kernels~\cite{Altarelli:1977zs}, explicitly
\begin{align}
&P_{bb}^+(z)=C_F\left[\frac{1+z^2}{(1-z)_{+}}+\frac{3}{2}\delta(1-z)\right],\nonumber\\
&P_{gb}^+(z)=C_F\left[\frac{1+(1-z)^2}{z}\right],\nonumber\\
&P_{bg}^+(z)=\frac{1}{2}[z^2+(1-z)^2],\nonumber\\
&P'_{bb}(z)=-C_{F}(1-z)+C_{F}\frac{1}{2}\delta(1-z),\nonumber\\
&P'_{bg}(z)=-z(1-z).
\end{align}
After replacing the bare PDF by the renormalized $\overline{\rm MS}$
PDF and integrating out the collinear region of the phase space
defined in the two-cutoff phase space slicing
method~\cite{Harris:2001sx}, the resulting sum of Eq.(\ref{e1}) and
the collinear part (the second term) of Eq. (\ref{e2}) yield the
remaining $\mathcal{O}(\alpha_{s})$ collinear contribution as:
\begin{eqnarray}
&& \sigma^{coll}= \int \hat{\sigma}^B\bigg[\frac{\alpha_s}{2\pi}
\frac{\Gamma(1-\epsilon)} {\Gamma(1-2\epsilon)}
\bigg(\frac{4\pi\mu^2_r}{s}\bigg)^\epsilon \bigg]
\{\tilde{G}_{b/p}(x_1,\mu_f) G_{\bar{b}/p}(x_2,\mu_f) +
G_{b/p}(x_1,\mu_f) \tilde{G}_{\bar{b}/p}(x_2,\mu_f) \nonumber
\\ && \hspace{1.2cm}
+\sum_{\alpha=b,\bar{b}}\bigg[\frac{A_1^{sc}(\alpha\rightarrow
\alpha g)}{\epsilon} +A_0^{sc}(\alpha\rightarrow \alpha
g)\bigg]G_{b/p}(x_1,\mu_f) G_{\bar{b}/p}(x_2,\mu_f) \nonumber
\\ && \hspace{1.2cm}
+(x_1\leftrightarrow x_2)\} dx_1dx_2-\frac{\alpha_s}{2\pi}C_F\sigma^B,
\end{eqnarray}
where
\begin{eqnarray}
&& A_1^{sc}(b\rightarrow bg)=A_1^{sc}(\bar{b}\rightarrow \bar{b}g)=C_F(2\ln\delta_s +3/2), \\
&& A_0^{sc}=A_1^{sc}\ln(\frac{s}{\mu_f^2}), \\
&&
\tilde{G}_{\alpha(=b,\bar{b})/p}(x,\mu_f)=\sum_{\beta=g,\alpha}\int_x^{1-
\delta_s\delta_{\alpha\beta}} \frac{dy}{y}
G_{\beta/p}(x/y,\mu_f)\tilde{P}_{\alpha\beta}(y)
\end{eqnarray}
with
\begin{eqnarray}
\tilde{P}_{\alpha\beta}(y)=P_{\alpha\beta}(y) \ln(\delta_c
\frac{1-y}{y} \frac{s}{\mu_f^2}) -P_{\alpha\beta}'(y).
\end{eqnarray}

The NLO total cross section for $pp\rightarrow H^-W^+$ in the
$\overline{\rm MS}$ factorization scheme is obtained by summing up
the Born, virtual, soft, collinear and hard noncollinear
contributions. In terms of the above notations, we have
\begin{eqnarray}
&& \sigma^{NLO}= \int dx_1dx_2 \{
\bigg[G_{b/p}(x_1,\mu_f)G_{\bar{b}/p}(x_2,\mu_f)+
(x_1\leftrightarrow x_2)\bigg](\hat{\sigma}^{B} +
\hat{\sigma}^{V}+ \hat{\sigma}^{S} +\hat{\sigma}^{\overline{HC}})\}
+\sigma^{coll} \nonumber
\\ && \hspace{0.4cm} +\sum_{(\alpha=g,\beta=b,\bar{b})}\int dx_1dx_2
\bigg[G_{\alpha/p}(x_1,\mu_f) G_{\beta/p}(x_2,\mu_f)
+(x_1\leftrightarrow x_2)\bigg]
\hat{\sigma}^{\overline{C}}(\alpha\beta\rightarrow H^-W^+ +X). \nonumber\\
\label{e3}
\end{eqnarray}
We note that the above expression contains no singularities, for
$2A_2^V +A_2^s =0$ and $2A_1^V +A_1^s +A_1^{sc}(b\rightarrow bg)
+A_1^{sc}(\bar{b}\rightarrow \bar{b}g) =0$. Namely, all the
$1/{\epsilon^2}$ and $1/{\epsilon}$ terms cancel in $\sigma^{NLO}$.
The apparent logarithmic $\delta_s$ and $\delta_c$ dependent terms
also cancel with the the hard noncollinear cross section
$\hat{\sigma}^{\overline{HC}}$ after numerically integrating over
its relevant phase space volume.
\section{MONTE CARLO SIMULATIONS}
Based on the work of Ref.~\cite{Eriksson:2006yt}, we discuss the
same signal in the mSUGRA scenario including the NLO QCD effects. In
the signal channel, $H^-$ decays leptonically, $H^- \to
\tau^{-}\bar{\nu}_{\tau}$ and $W^+$ decays hadronically, $W^+ \to
q\bar{q}'(q=u,c,q'=d,s)$. For simplicity, we only consider hadronic
decays of the $\tau$ lepton, $\tau \to \nu_{\tau}+hadrons$. The
resulting signature is $2j+\tau_{jet}+\not{\!p}_{T}$, where the
missing transverse momentum $\not{\!p}_{T}$ is carried away by the
two neutrinos, $\tau_{jet}$ comes from $\tau$ decay and $2j$ come
from $W$ boson decay. The transverse mass is defined as:
\begin{equation}
 m_{\bot}=\sqrt{2p_{T,\tau jet}\not{\!p}_{T}[1-\cos(\Delta \phi)]},
\end{equation}
where $\Delta \phi$ is the azimuthal angle between $p_{T,\tau jet}$
and $\not{\!p}_{T}$. As pointed out in Ref.~\cite{Eriksson:2006yt},
the $m_{\bot}$ distribution will have a peak with the upper edge of
the peak given by the mass of the charged Higgs boson. The two light
jets can be distinguished by calling them hard (with momentum
$p_{hj}$) and soft (with momentum $p_{sj}$) according to the larger
and smaller value of their transverse momentum $p_T$, respectively.
Our study is performed at parton level, without considering parton
showering or hadronization, and the detector effects also not be
considered. Event generation is performed with help of PYTHIA
v6.206~\cite{Sjostrand:2001yu} and TAUOLA
v2.7~\cite{Jadach:1990mz,Golonka:2003xt} is used to perform the
decay of $\tau$ lepton.

The cuts we have used are shown in Table \ref{ad}, which are
the same as in Ref.~\cite{Eriksson:2006yt} in order to compare our results with theirs.
\begin{table}
\begin{tabular}{|c|c|}
\hline
Basic cuts & Additional cuts [all in GeV] \\
\hline
$|\eta_{\tau_{jet}}|<2.5$ & $p_{T,\tau_{jet}}>50,\not{\!p}_T >50$ \\
$|\eta_j|<2.5$     &   $70<m_{jj}<90$  \\
$\Delta R_{jj}>0.4$  &   $m_{\bot}>100$  \\
$\Delta R_{\tau_{jet}j}>0.5$  & $p_{T,hj}>50,p_{T,sj}>25$ \\
$p_{T,jet}>20{\rm GeV}$ & \\
\hline
\end{tabular}\caption{Basic cuts and additional cuts used} \label{ad}
\end{table}
Here the basic cuts define a signal region that corresponds to the
sensitive region of a real detector and the additional cuts are used
to suppress both background and detector misidentifications. The
dominant irreducible SM background for our signature
$2j+\tau_{jet}+\not{\!p}_T$ comes from $W+2j$ production with
$W \to \tau \nu _{\tau}$. We use
ALPGEN~\cite{Mangano:2002ea} to repeat the background calculations
of Ref.\cite{Eriksson:2006yt}, and the same results can be obtained.
The $W^+ +2j$ background mainly comes from $ug$ and $\bar{d}g$
initial states, while $W^- +2j$ background is mainly due to $dg$ and
$\bar{u}g$ initial states. Detailed descriptions about the
backgrounds and cuts can be found in Ref.\cite{Eriksson:2006yt}, and
our simulation results will be discussed below.

\section{NUMERICAL RESULTS}
The arrangement of this part is as follow. First, we present the NLO
QCD calculations of both total cross sections and differential cross
sections. Then we turn to the simulation results under several
groups of cuts and mSUGRA parameters.
\subsection{NLO cross section calculations}
In the numerical calculations, we used the following set
of SM parameters~\cite{Yao:2006px}:
\begin{align}
 &\alpha _{ew}(m_{W})=\frac{e^2}{4\pi}=1/128,m_{W}=80.40{\rm GeV},
 m_{Z}=91.1876{\rm GeV},\nonumber\\
 &m_{t}=174.2{\rm GeV},m_{b}(m_{b})=4.2{\rm GeV},\alpha _{s}(M_{Z})=0.1176,
 s_W^2=0.23122.
\end{align}
The running QCD coupling $\alpha_s(Q)$ is evaluated at the two-loop
order~\cite{Gorishnii:1990zu} and the CTEQ6M
PDF~\cite{Pumplin:2002vw} is used throughout this paper to calculate
various cross sections, either at the LO or the NLO. As for the
factorization and renormalization scales, we always choose
$\mu_{f}=m_{{\rm av}}=(m_{H^{\pm}}+m_{W})/2$ and $\mu_{r}=\mu_{f}$,
unless specified otherwise. Moreover, as to the Yukawa couplings of
the bottom quark and top quark, we took the running masses $m_b(Q)$
and $m_t(Q)$ evaluated by the NLO formula~\cite{Carena:1999py}:
\begin{align}
&m_b(Q)=U_6(Q,m_t)U_5(m_t,m_b)m_b(m_b),\nonumber\\
&m_t(Q)=U_6(Q,m_t)m_t(m_t),
\end{align}
with $m_{b}(m_{b})=4.2{\rm GeV}$~\cite{Yao:2006px}. The evolution
factor $U_f$ is
\begin{eqnarray}
U_f(Q_2,Q_1)=\bigg(\frac{\alpha_s(Q_2)}{\alpha_s(Q_1)}\bigg)^{d^{(f)}}
\bigg[1+\frac{\alpha_s(Q_1)-\alpha_s(Q_2)}{4\pi}J^{(f)}\bigg], \nonumber \\
d^{(f)}=\frac{12}{33-2f}, \hspace{1.0cm}
J^{(f)}=-\frac{8982-504f+40f^2}{3(33-2f)^2} \, ,
\end{eqnarray}
where $f$ is the number of the active light quarks. We use both the
$\overline{\rm MS}$ and the OS renormalization scheme for top quark
in our calculations and find good agreement in these two schemes. We
will only show the numerically results in the $\overline{\rm MS}$
scheme unless specified otherwise.

The values of the MSSM parameters taken in our numerical
calculations were constrained within mSUGRA, in which there are only
five free input parameters at the grand unification (GUT) scale.
They are $m_{1/2}$, $m_0$, $A_{0}$, $\tan \beta$, and the sign of
$\mu$, where $m_{1/2}$, $m_{0}$, $A_0$, $\mu$ are, respectively, the
universal gaugino mass, scalar mass, the trilinear soft breaking
parameter, and the Higgs superfield mass term in the superpotential.
Given those parameters, all the
MSSM parameters at the weak scale are determined in the mSUGRA
scenario by using the program package SPHENO~\cite{Porod:2003um}.

\begin{figure}[!ht]
\includegraphics[width=0.90\textwidth]{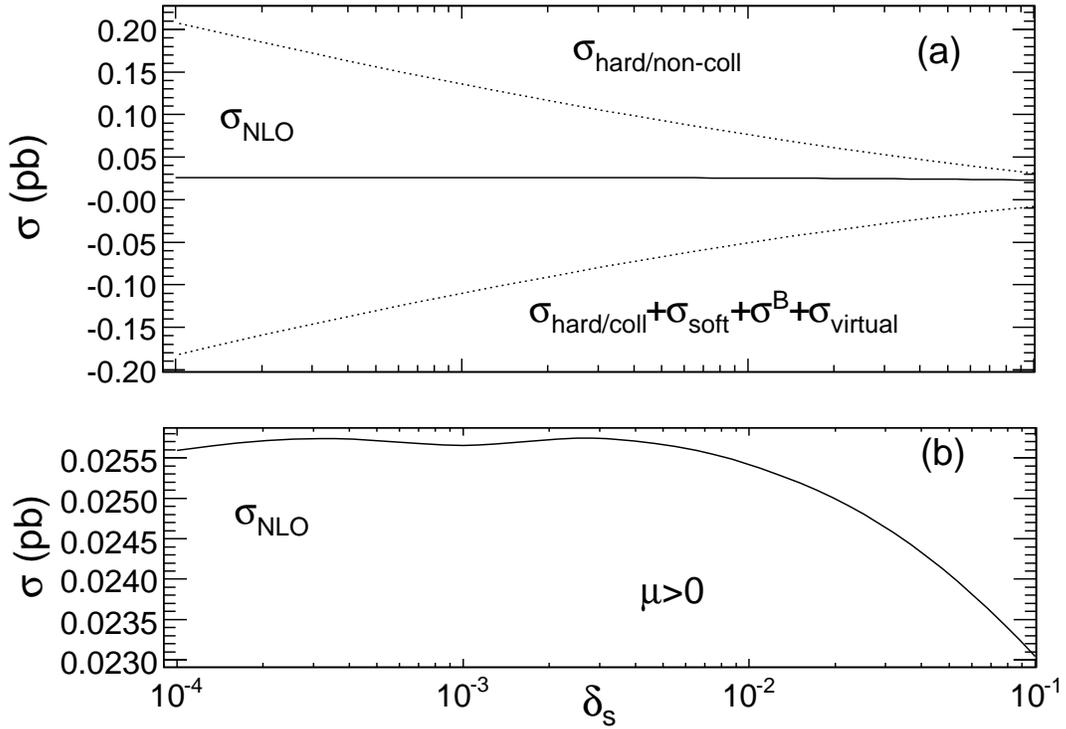}
\caption[]{Dependence of the NLO total cross sections for the
$H^-W^+$ production at the LHC on the theoretical cutoff scale
$\delta_s$ with $\delta_c=\delta_s/100$, assuming the mSUGRA model
with $m_0=150\ {\rm GeV},m_{1/2}=300\ {\rm GeV},A_0=300\ {\rm
GeV},\tan\beta=40$ and $\mu>0$.} \label{f1}
\end{figure}

In Fig.~\ref{f1}, we show the dependence of the NLO QCD predictions
on the two arbitrary theoretical cutoff scales $\delta_s$ and
$\delta_c$, introduced in the two-cutoff phase space slicing method,
where we have set $\delta_c=\delta_s/100$ to simplify the study. The
NLO total cross section can be separated into two classes of
contributions. One is the $2 \to 2$ rate contributed by the Born
level, and the ${\cal{O}}(\alpha_s)$ virtual, soft and hard
collinear real emission corrections, denoted as $\hat{\sigma}^{B}$,
$\hat{\sigma}^{V}$, $\hat{\sigma}^{S}$, and $\sigma^{coll}$ in
Eq.(\ref{e3}). Another is the $2 \to 3$ rate contributed by the
${\cal{O}}(\alpha_s)$ hard noncollinear real emission corrections,
denoted as $\hat{\sigma}^{\overline{HC}}$ and
$\hat{\sigma}^{\overline{C}}$ in Eq.(\ref{e3}). As noted in the
previous section, the $2 \to 2$ and $2 \to 3$ rates depend
individually on $\delta_s$ and $\delta_c$, but their sum should not
depend on any of the theoretical cutoff scales. This is clearly
illustrated in Fig.~\ref{f1}, where $\sigma_{NLO}$ is almost
unchanged for $\delta_s$ between $ 10^{-4}$ and $10^{-2}$, and is
about 25.6\,fb. Therefore, we take $\delta_s=10^{-3}$ and
$\delta_c=\delta_s/100$ in the numerical calculations below.

\begin{figure}[!ht]
\includegraphics[width=0.90\textwidth]{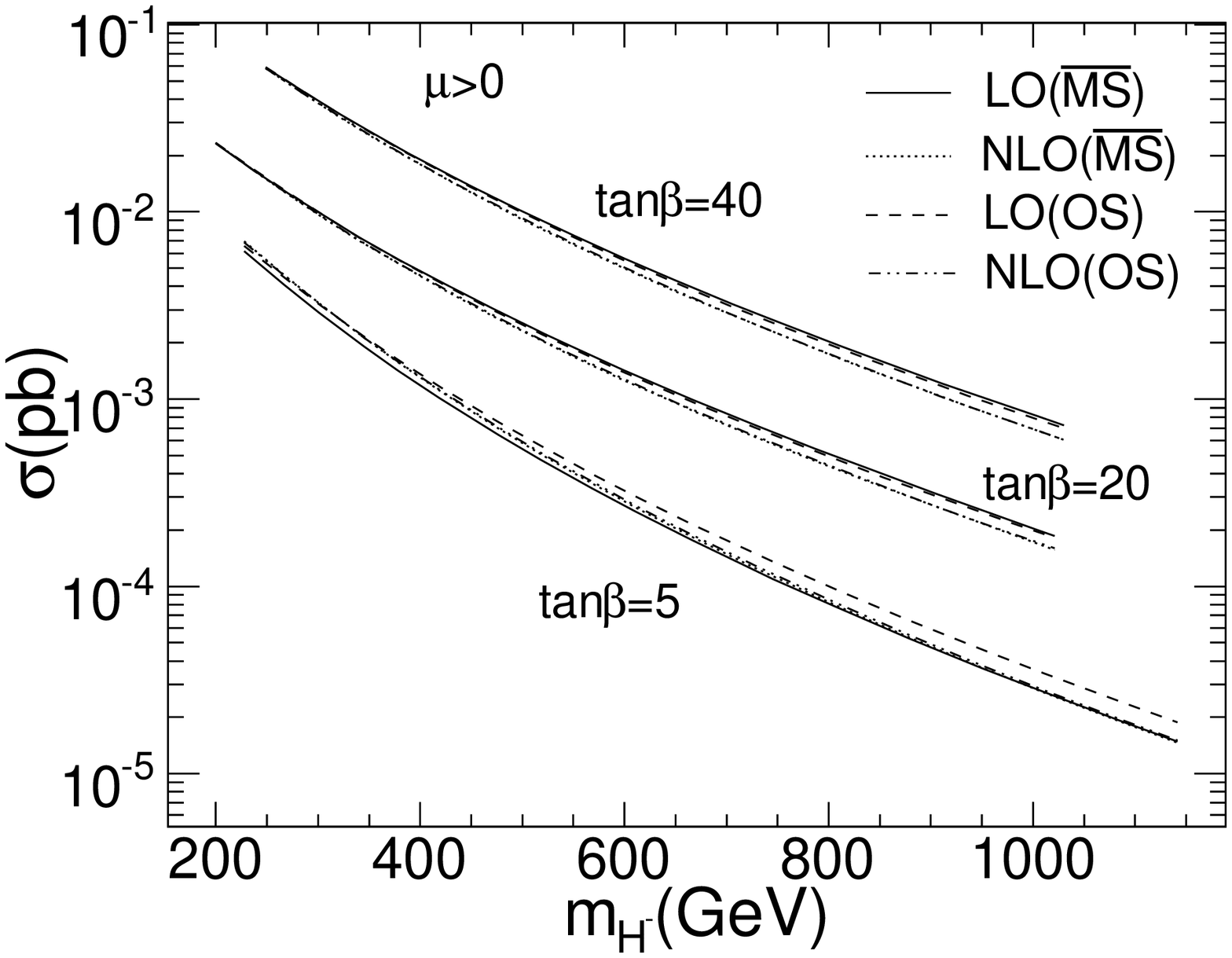}
\caption[]{Dependence of the total cross sections for the $H^-W^+$
production at the LHC on $m_{H^-}$, assuming $m_0=150\ {\rm GeV}$, and
$A_0=300\ {\rm GeV}$.}\label{f2}
\end{figure}

Fig.~\ref{f2} shows the total cross sections for $pp \to H^-W^+$ at
the LHC in both the $\overline{\rm MS}$ scheme and the OS scheme as
a function of $m_{H^{-}}$ for $\tan \beta=$ 5,\,20 and 40,
respectively, assuming $m_{0}=150 {\rm GeV}$, and $A_0=300 {\rm
GeV}$. The results in the two schemes are almost the same. The total
cross sections decrease with the increasing $m_{H^-}$. In general,
the NLO QCD corrections enhance the total cross sections for small
$\tan \beta$, but reduce for large $\tan \beta$.

\begin{figure}[!ht]
\includegraphics[width=0.90\textwidth]{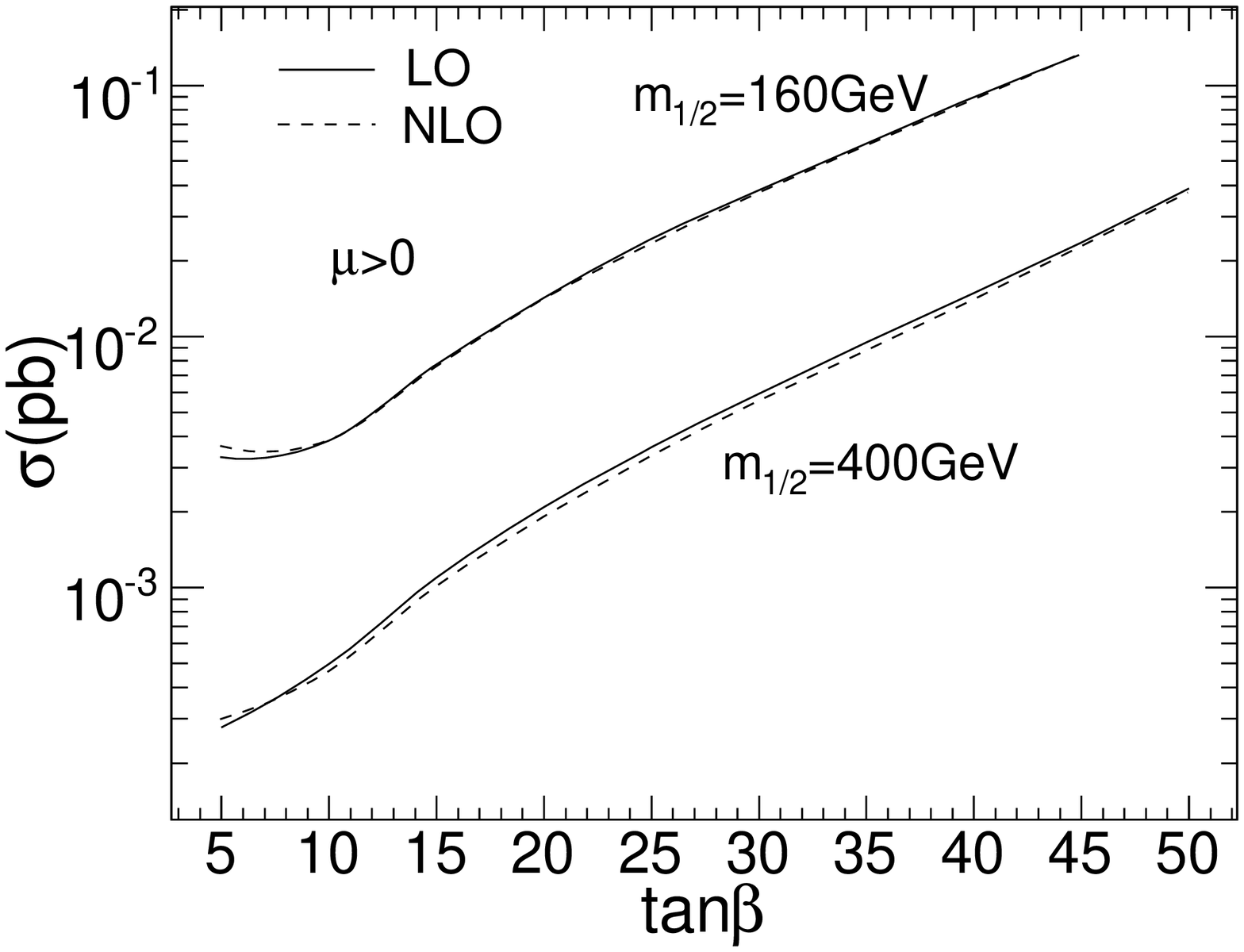}
\caption[]{Dependence of the total cross sections for the $H^-W^+$
production at the LHC on $\tan\beta$, assuming $m_0=150\ {\rm GeV}$,
and $A_0=300\ {\rm GeV}$.} \label{f4}
\end{figure}

In Fig.~\ref{f4}, the total cross sections for $pp \to H^-W^+$ at the
LHC are plotted as a function of $\tan \beta$ for two representative
values of $m_{1/2}$. When $\tan \beta$ ranges between 5 and 45,
$m_{H^-}$ varies from $290 \ \rm{GeV}$ to $185 \ \rm{GeV}$, and from
$595\ \rm{GeV}$ to $402 \ \rm{GeV}$ for $m_{1/2}=160 \ {\rm GeV}$
and $400\ \rm{GeV}$, respectively. From Fig.~\ref{f4} we can clearly
see that the total cross sections increase with the increasing $\tan
\beta$ and decrease with the increasing $m_{1/2}$. For large $\tan
\beta(>40)$ and $m_{1/2}=160\ {\rm GeV}$, the LO and the NLO total
cross sections can be over 100 fb.

\begin{figure}[!ht]
\includegraphics[width=0.90\textwidth]{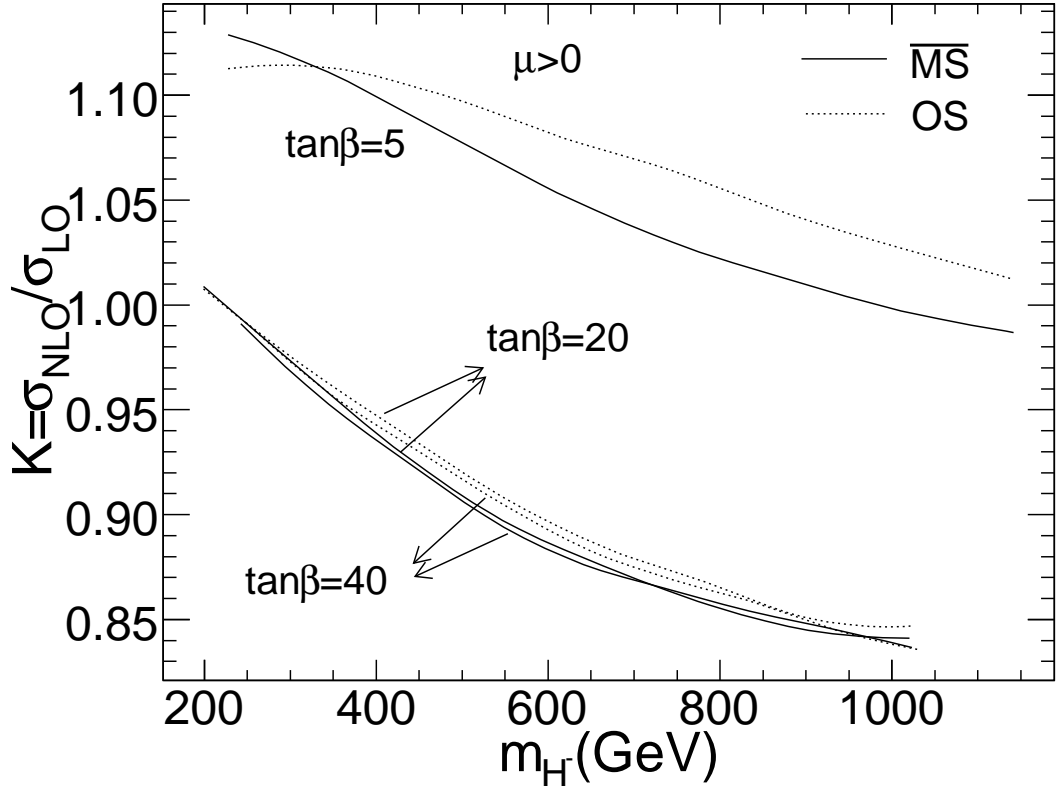}
\caption[]{Dependence of the K factor for the $H^-W^+$ production at
the LHC on $m_{H^-}$, assuming $m_{0}=150\ {\rm GeV}$, and
$A_0=300\ {\rm GeV}$.} \label{f3}
\end{figure}

Fig.~\ref{f3} gives the dependence of the K factor (defined as the
ratio of the NLO total cross sections to the LO ones in the
$\overline{\rm MS}$ scheme) on $m_{H^-}$ for
$H^-W^+$ production, based on the results in Fig.~\ref{f2}. It can be seen that
the results in the two schemes are in good agreement.
For instance, the difference of the K factors in the two schemes is within 4\% for
$\tan \beta=5$ and less than 2\% for $\tan \beta=20$ and 40.
In general, the K factor decreases with the increasing $m_{H^-}$.
For $\tan \beta=5$, the K factors can increase to 1.1 when $m_{H^-}<
400\ \rm{GeV}$. While for $\tan \beta =20$ and 40, the K factors
decrease below 0.9 when $m_{H^-} > 500 \ \rm{GeV}$.

\begin{figure}[!ht]
\includegraphics[width=0.80\textwidth]{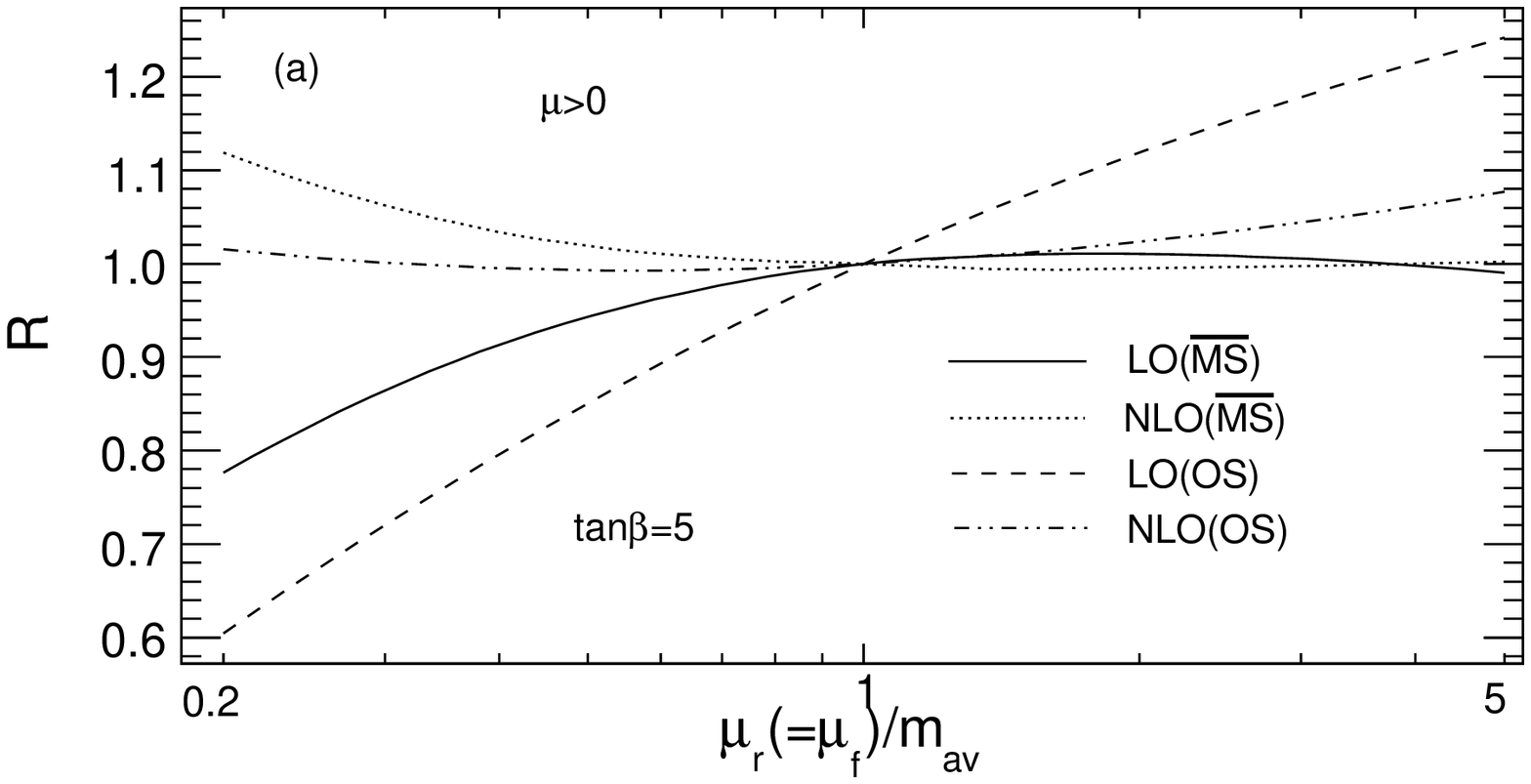}
\includegraphics[width=0.80\textwidth]{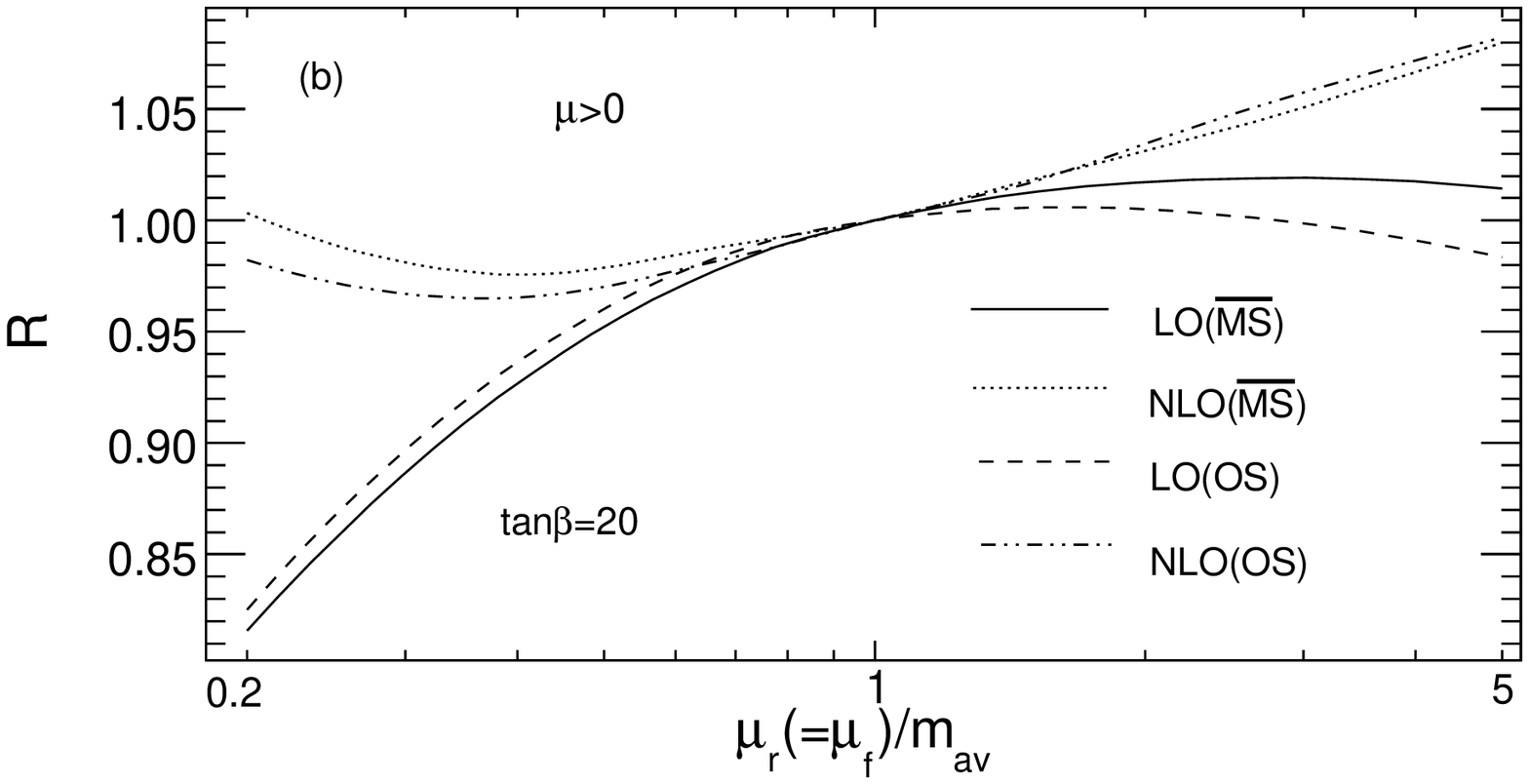}
\includegraphics[width=0.80\textwidth]{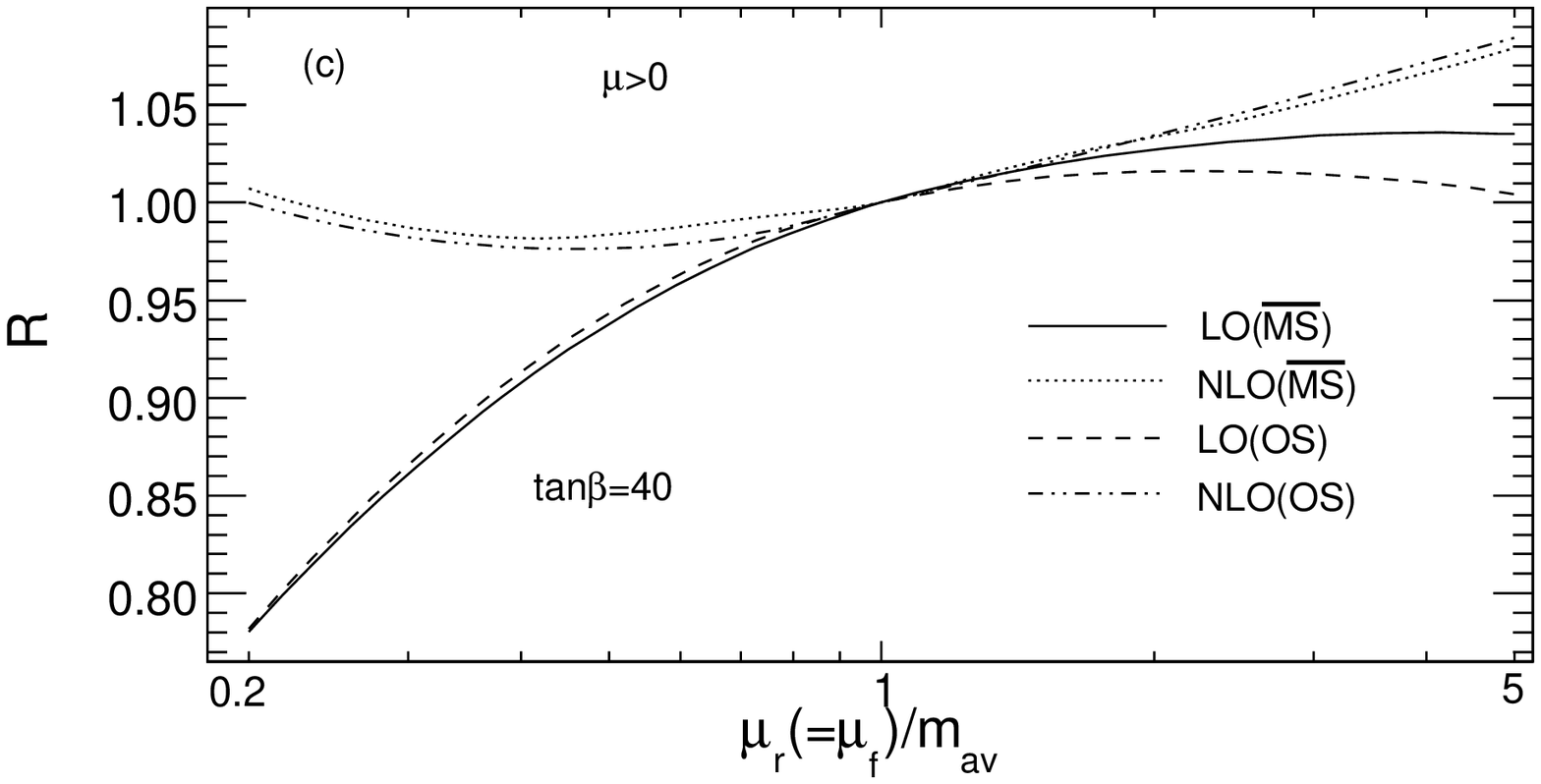}
\caption[]{Dependence of the ratio R on the
renormalization/factorization scale ($\mu_r=\mu_f$) for the $H^-W^+$
production at the LHC, assuming: (a) $m_0=150\ {\rm GeV},
m_{1/2}=170\ {\rm GeV},A_0=300\ {\rm GeV}$, and $\tan \beta=5$; (b)
$m_0=150\ {\rm GeV},m_{1/2}=300\ {\rm GeV},A_0=300\ {\rm GeV}$, and
$\tan\beta=20$; (c) $m_0=150\ {\rm GeV},m_{1/2}=300\ {\rm
GeV},A_0=300\ {\rm GeV}$, and $\tan\beta=40$.} \label{f6}
\end{figure}

Figs.~\ref{f6} shows the dependence of the total cross
sections for $pp \to H^-W^+$ production at the LHC on the
renormalization scale ($\mu_r$) and the factorization scale
($\mu_f$), with $\mu_r=\mu_f$. We defined R as the ratio of
the cross sections (LO, NLO) to their values at central scale,
$\mu _r=\mu _f=m_{av}=(m_{H^{\pm}}+m_W)/2$,
always assuming $\mu _r=\mu _f$ for simplicity.
For three values of $\tan \beta$, the scale
dependence of the NLO total cross sections reduced when going from LO to
NLO in both the $\overline{\rm MS}$ and the OS scheme.
For example, in the $\overline{\rm MS}$ scheme, the ratio R at the LO
vary from 0.78 to 1.03 when
$\mu_r=\mu_f$ ranges between $0.2m_{{\rm av}}$ and $5m_{{\rm av}}$,
while the NLO ones vary from 0.98 to
1.08, for $\tan \beta=40$.

\begin{figure}[!ht]
\includegraphics[width=0.80\textwidth]{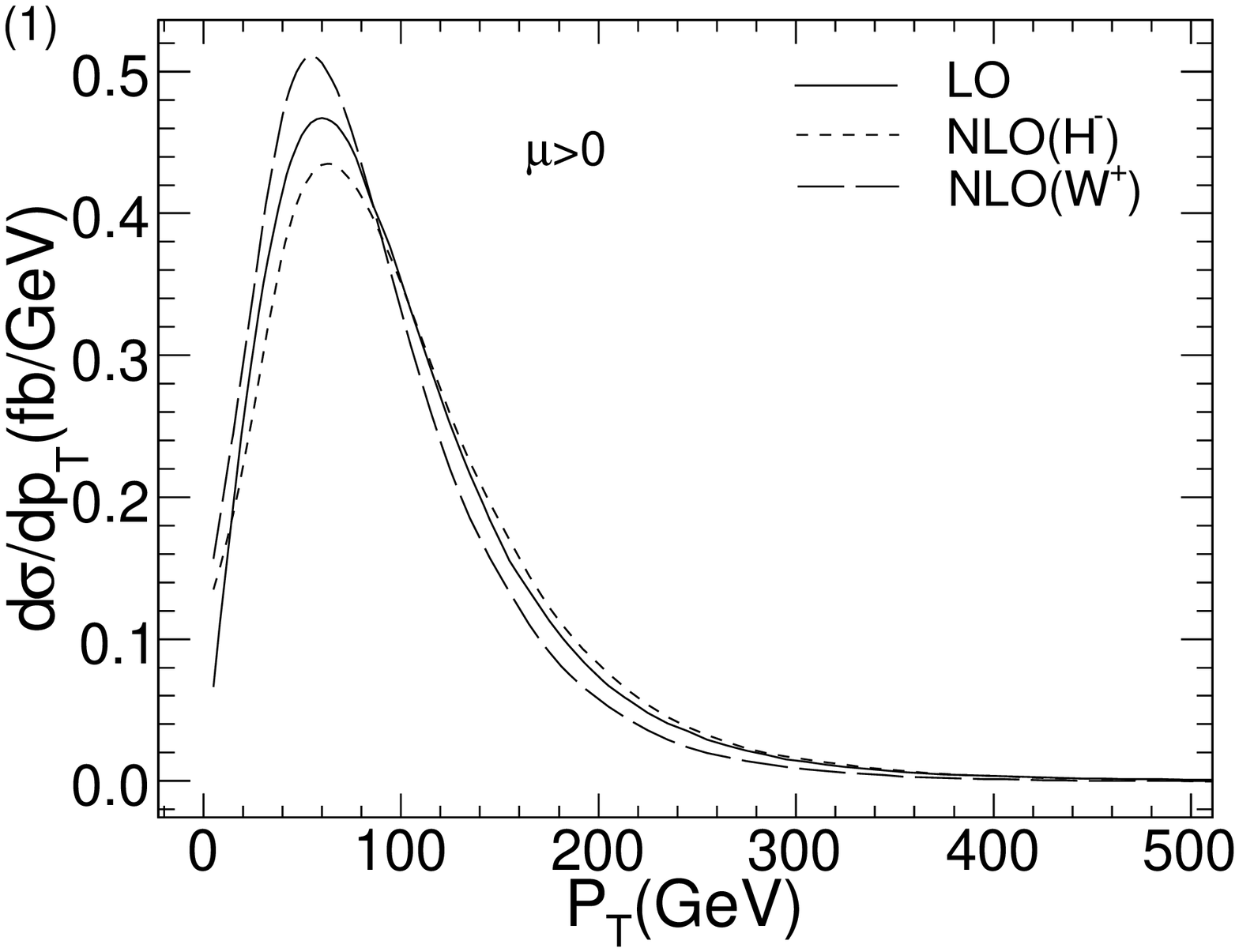}
\includegraphics[width=0.80\textwidth]{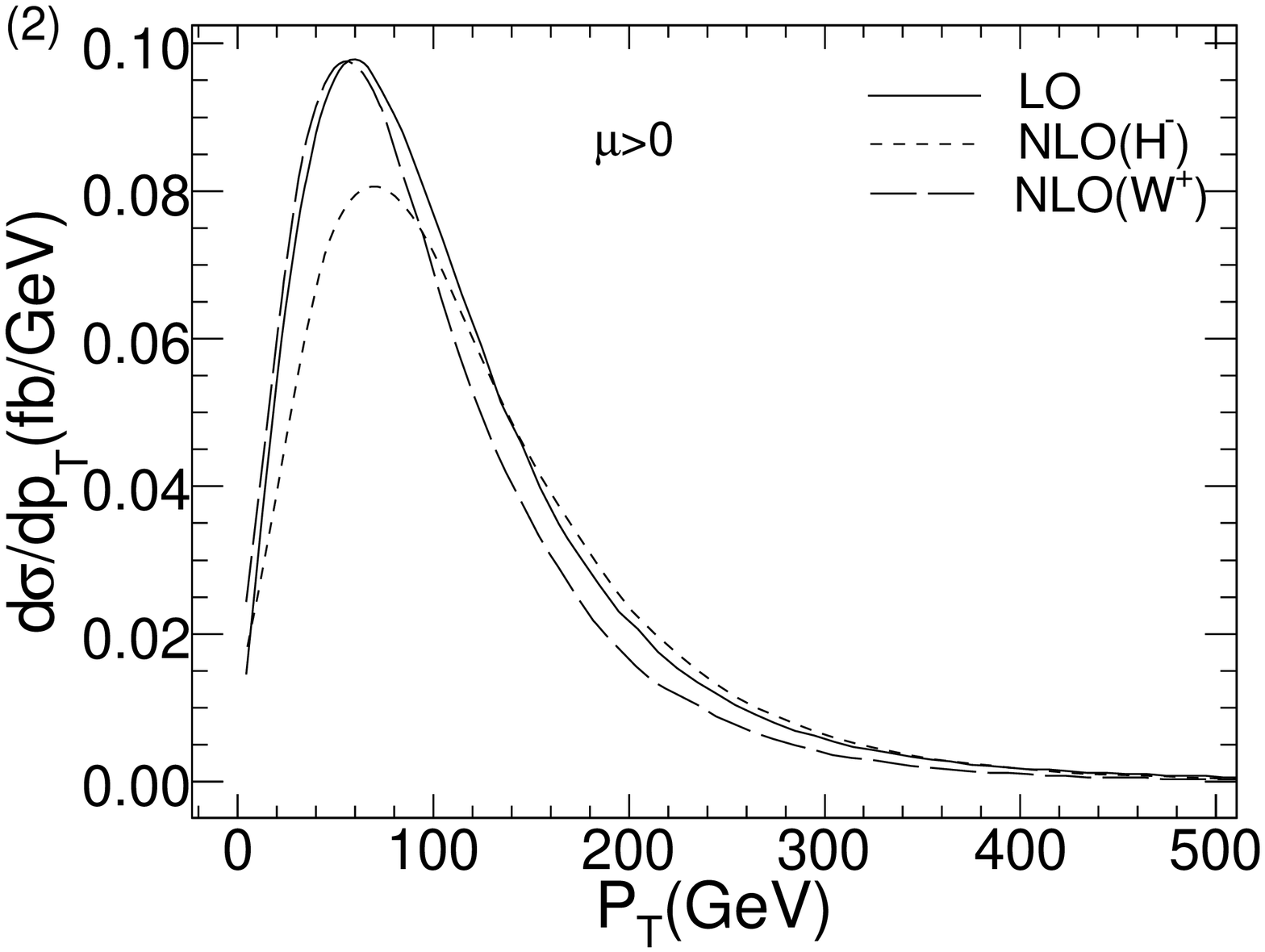}
\caption[]{Differential cross sections in the transverse momentum
($p_{T}$) of the $H^-$ and the $W^+$ bosons, for the $H^-W^+$
production at the LHC, assuming: (1) $m_0=200\ {\rm
GeV},m_{1/2}=180\ {\rm GeV},A_{0}=250\ {\rm GeV}$, and $\tan\beta=40$;
(2) $m_0=150\ {\rm GeV},m_{1/2}=400\ {\rm GeV},A_{0}=300\ {\rm GeV}$, and
$\tan\beta=40$.} \label{f9}
\end{figure}

Fig.~\ref{f9} shows the differential cross sections as the functions
of the transverse momentum $p_{T}$ of the $H^-$ and the $W^+$ in the
associated production of the $H^-W^+$ pairs at the LHC. In case (1),
the NLO QCD corrections can enhance and reduce the differential
cross sections in the medium $p_T$ region of the $W^+$ and the
$H^-$, respectively, and are negligible small in both the high and
the low $p_T$ region. In case (2), the NLO QCD corrections reduce
the differential cross section significantly in the medium $p_{T}$
region of the $H^-$, otherwise the NLO QCD corrections can be
neglected.

\begin{figure}[!ht]
\includegraphics[width=0.80\textwidth]{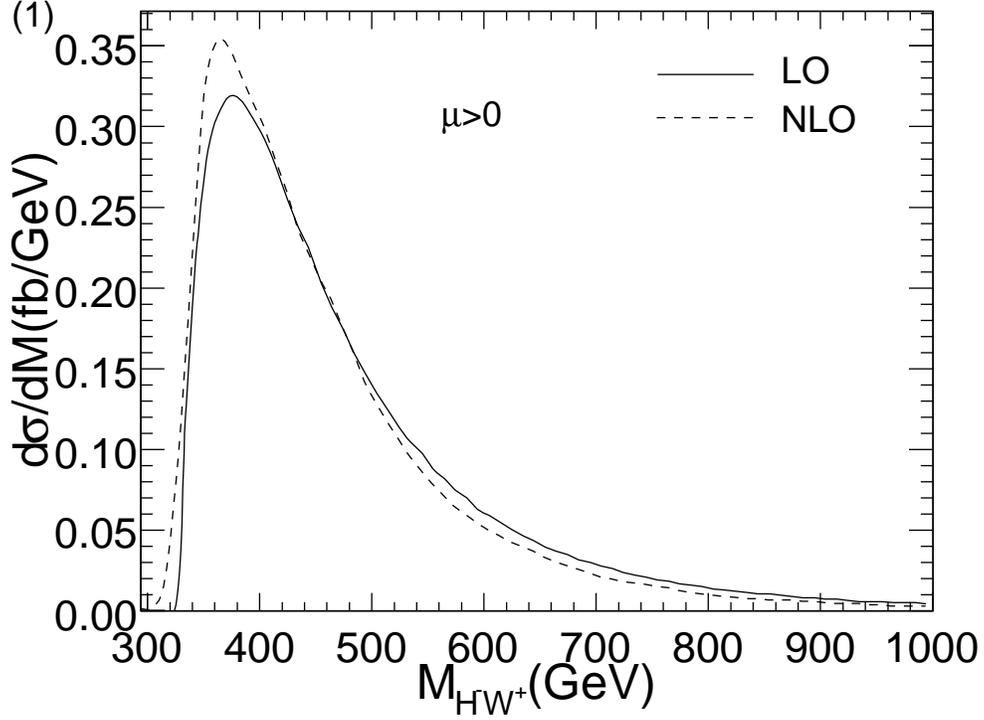}
\includegraphics[width=0.80\textwidth]{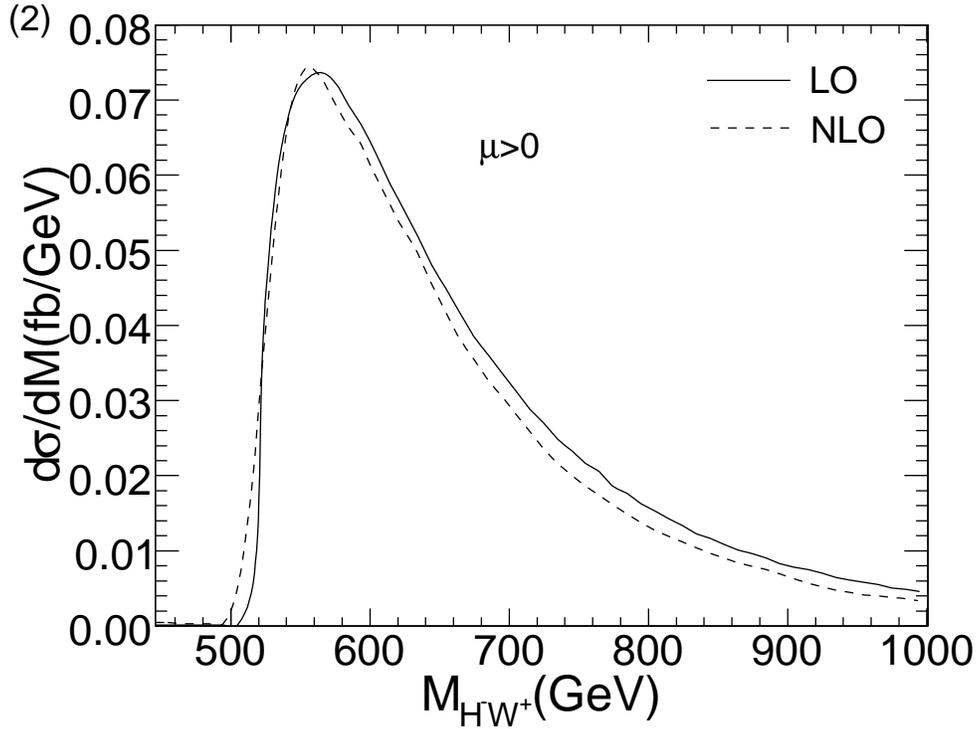}
\caption[]{Differential cross sections in the invariant mass
($M_{H^-W^+}$), for the $H^-W^+$ production at the LHC, assuming:
(1) $m_0=200\ {\rm GeV},m_{1/2}=180\ {\rm GeV},A_{0}=250\ {\rm GeV}$, and
$\tan\beta=40$; (2) $m_0=150\ {\rm GeV},m_{1/2}=400\ {\rm
GeV},A_{0}=300\ {\rm GeV}$, and $\tan\beta=40$.} \label{f10}
\end{figure}

In Fig.~\ref{f10} we display the differential cross sections as the
functions of the invariant mass $M_{H^-W^+}$ of the $H^-W^+$ pairs
produced at the LHC. In case (1), the NLO QCD corrections enhance
the LO differential cross sections more, which can reach 10\%, in
the medium values of the $M_{H^-W^+}$, but are negligible small in
both the high and the low values of the $M_{H^-W^+}$. In case (2),
the NLO QCD corrections reduce the LO differential cross sections in
the high values of the $M_{H^-W^+}$, while the corrections are
relatively small in the low values of the $M_{H^-W^+}$.

Note that our numerical results of the NLO QCD corrections to the
total cross sections are different from the ones given in
Ref.~\cite{Hollik:2001hy}, where the corrections are always negative
and the magnitude can reach 30\%. We also used the same parameters
as in Ref.~\cite{Hollik:2001hy} to compare with their results, but
our results are still different from theirs.

\subsection{Simulation results}
\begin{figure}[!ht]
\includegraphics[width=0.80\textwidth]{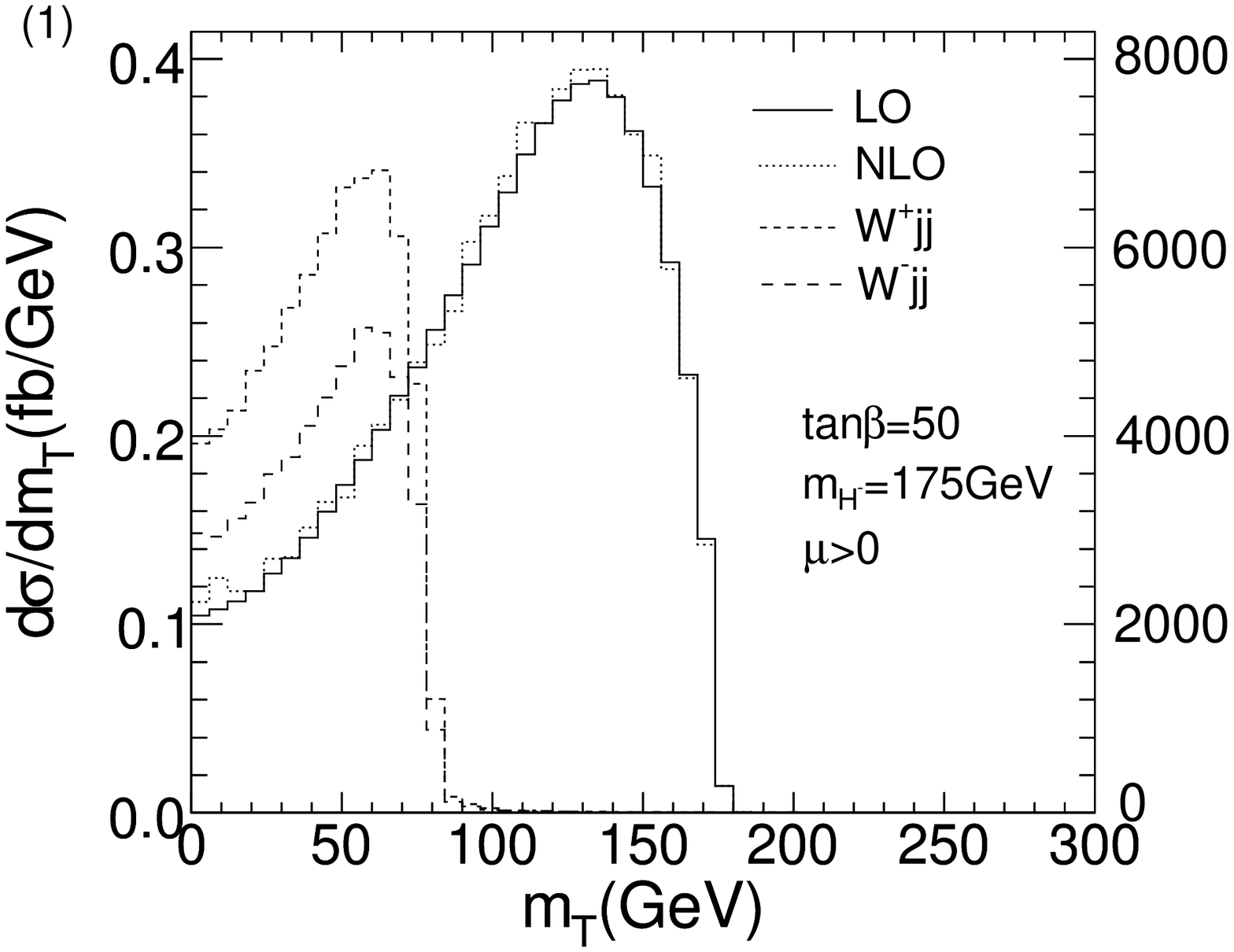}
\includegraphics[width=0.80\textwidth]{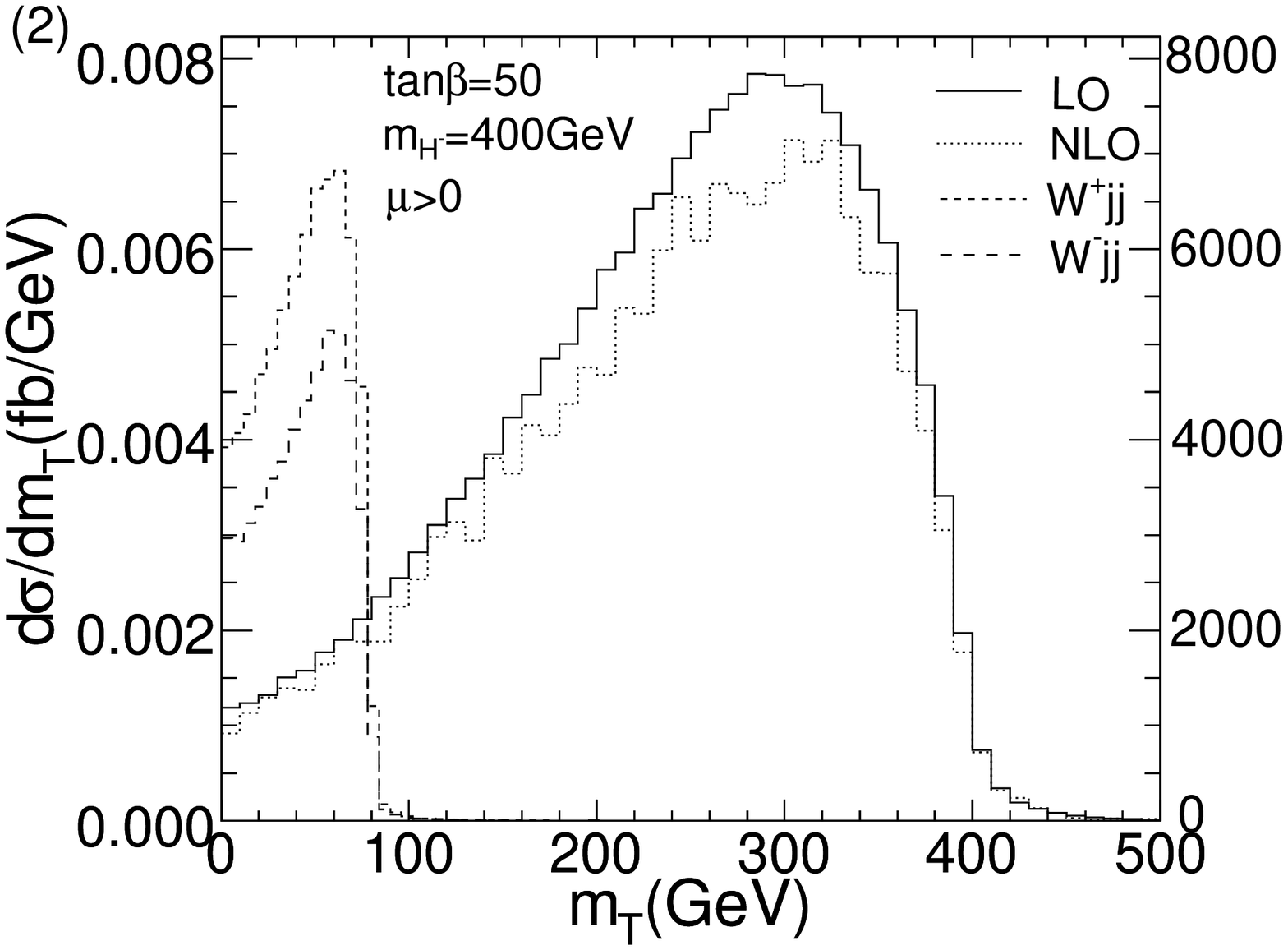}
\caption[]{$m_T$ distributions for the signal and backgrounds after
basic cuts, assuming: (1) $m_0=200\ {\rm GeV},m_{1/2}=147\ {\rm GeV},$
and $A_{0}=200\ {\rm GeV}$; (2) $m_0=320\ {\rm GeV},m_{1/2}=400\ {\rm
GeV}$, and $A_{0}=\ 300{\rm GeV}$. The left axis scale is for the cross section
of the signal, while the right one is for the backgrounds.} \label{f11}
\end{figure}

\begin{figure}[!ht]
\includegraphics[width=0.80\textwidth]{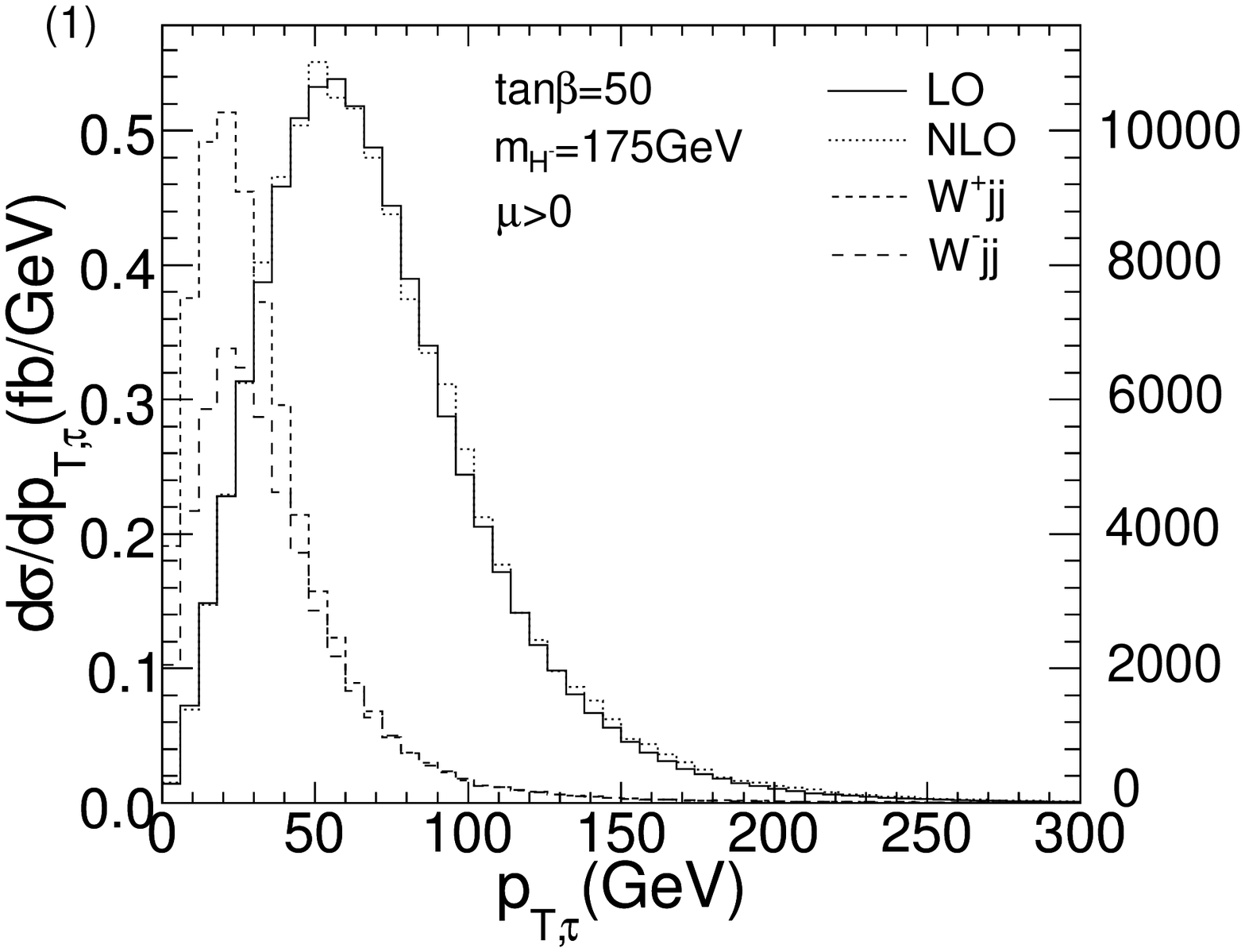}
\includegraphics[width=0.80\textwidth]{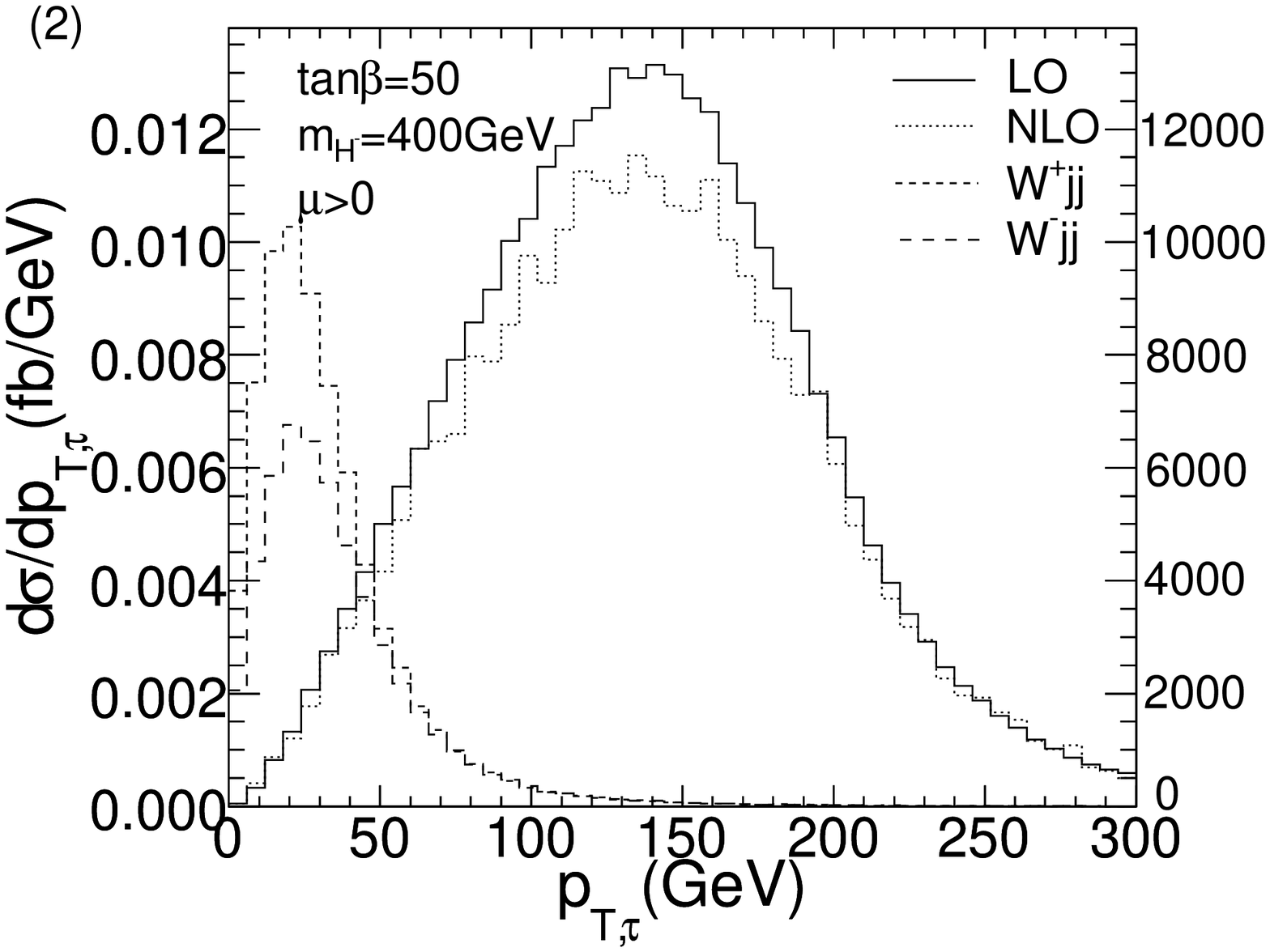}
\caption[]{Transverse momentum distributions of the $\tau$ jet for the
signal and backgrounds after basic cuts, assuming: (1) $m_0=\ 200{\rm
GeV},m_{1/2}=147\ {\rm GeV}$, and $A_{0}=200\ {\rm GeV}$; (2)
$m_0=320\ {\rm GeV},m_{1/2}=400\ {\rm GeV}$, and $A_{0}=300\ {\rm GeV}$.
The left axis scale is for the cross section
of the signal, while the right one is for the backgrounds.}
\label{f12}
\end{figure}

\begin{figure}[!ht]
\includegraphics[width=0.80\textwidth]{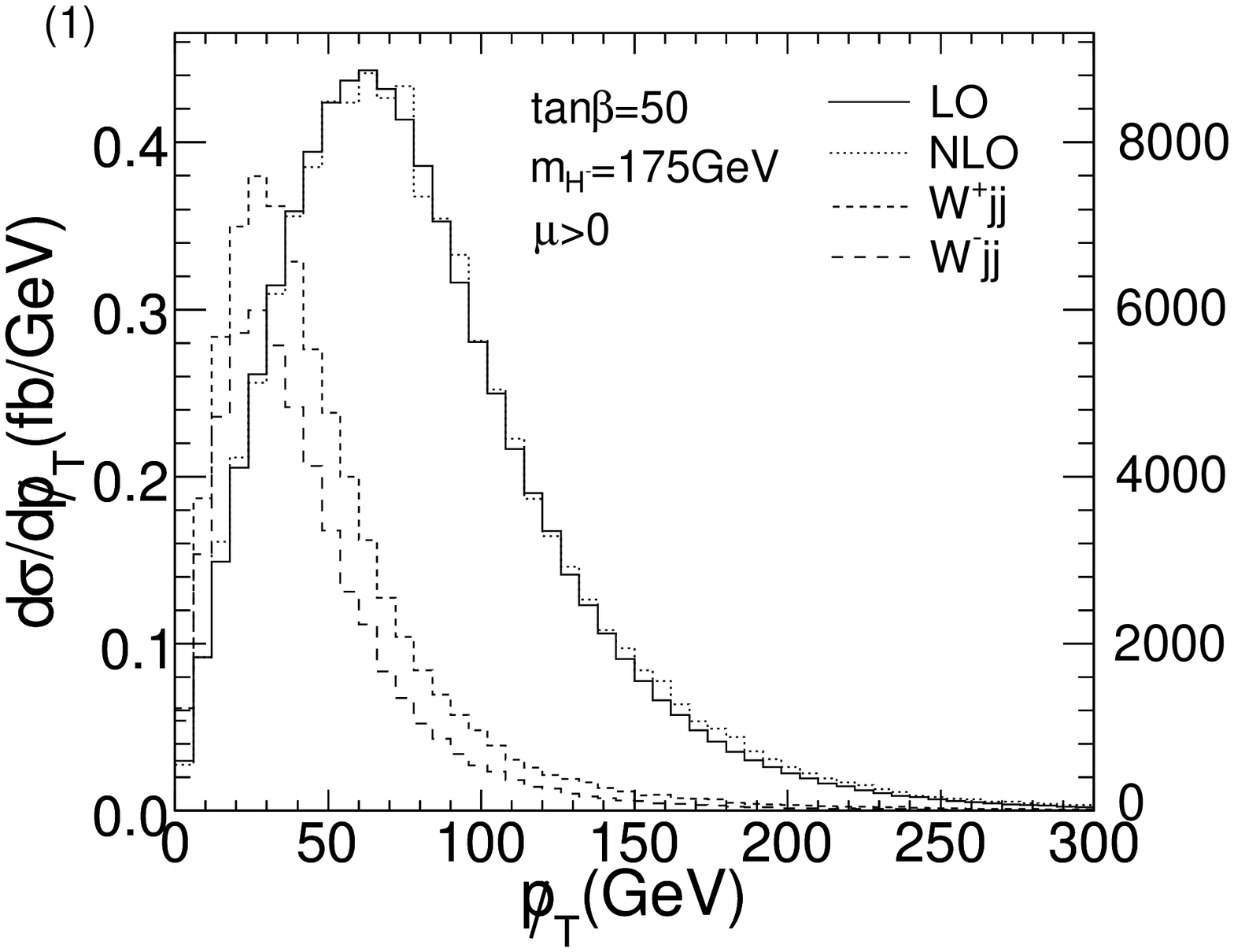}
\includegraphics[width=0.80\textwidth]{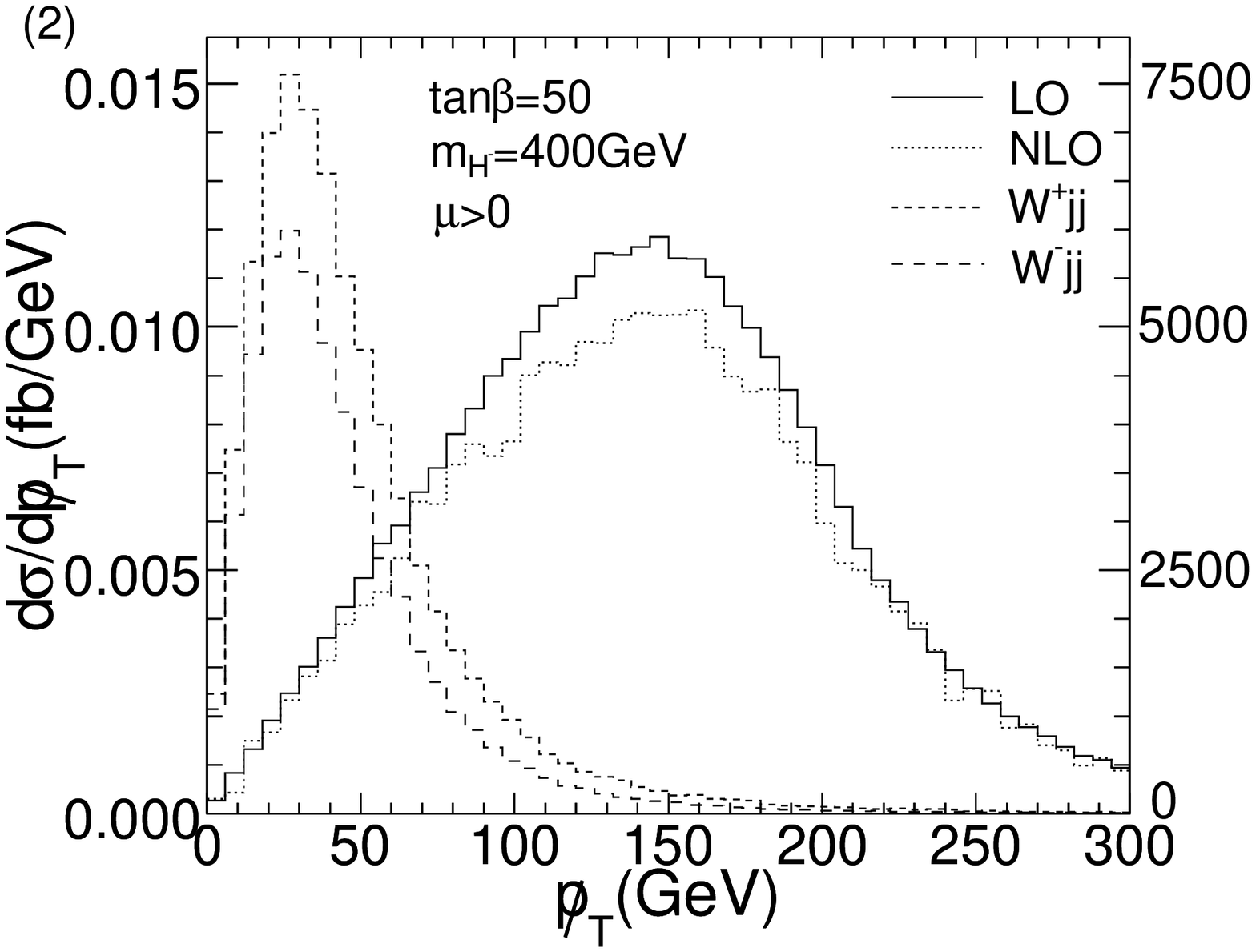}
\caption[]{Missing transverse momentum distributions for the signal
and backgrounds after basic cuts, assuming: (1) $m_0=200\ {\rm
GeV},m_{1/2}=147\ {\rm GeV}$, and $A_{0}=200\ {\rm GeV}$; (2)
$m_0=320\ {\rm GeV},m_{1/2}=400\ {\rm GeV}$, and $A_{0}=300\ {\rm GeV}$.
The left axis scale is for the cross section
of the signal, while the right one is for the backgrounds.}
\label{f13}
\end{figure}

\begin{figure}[!ht]
\includegraphics[width=0.80\textwidth]{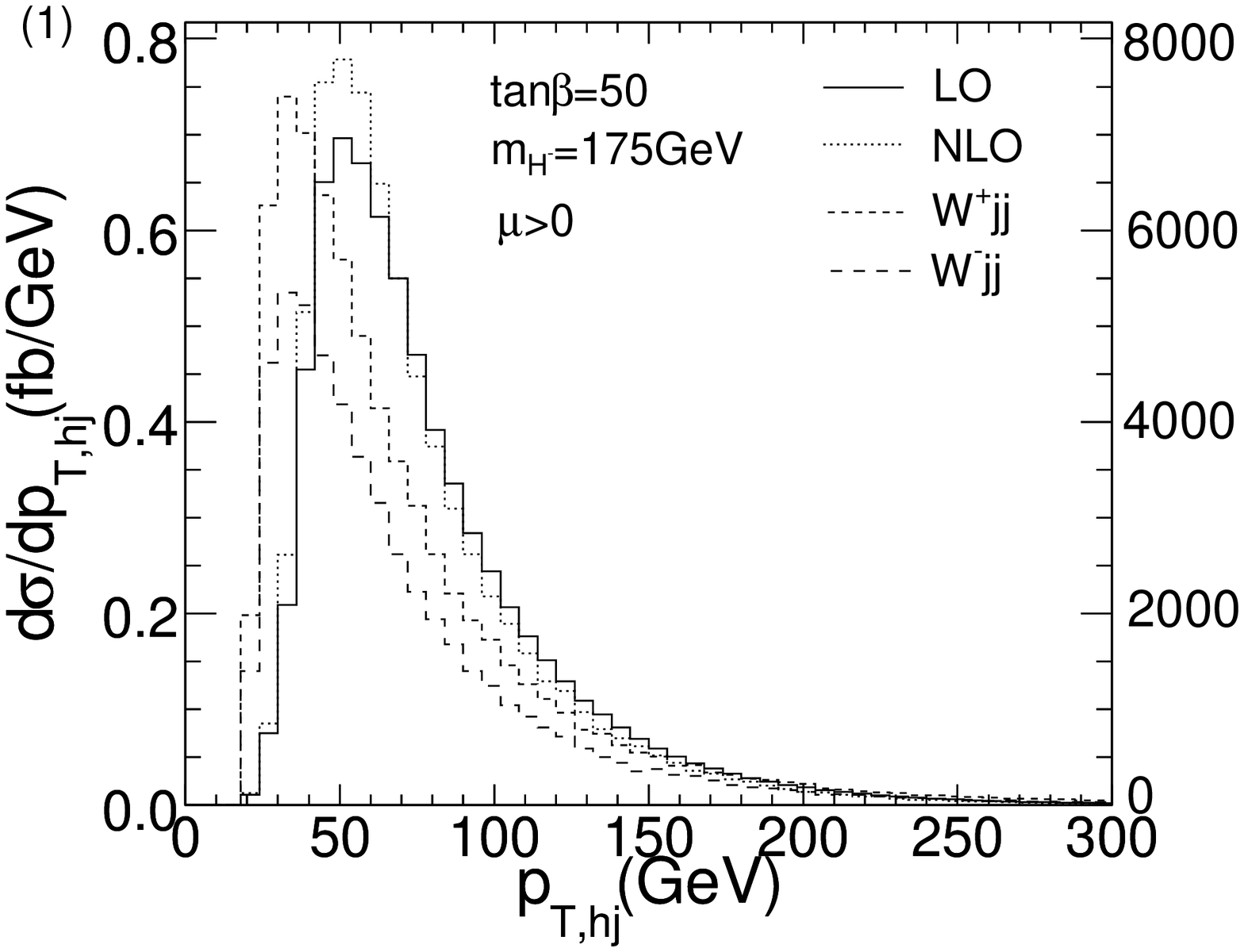}
\includegraphics[width=0.80\textwidth]{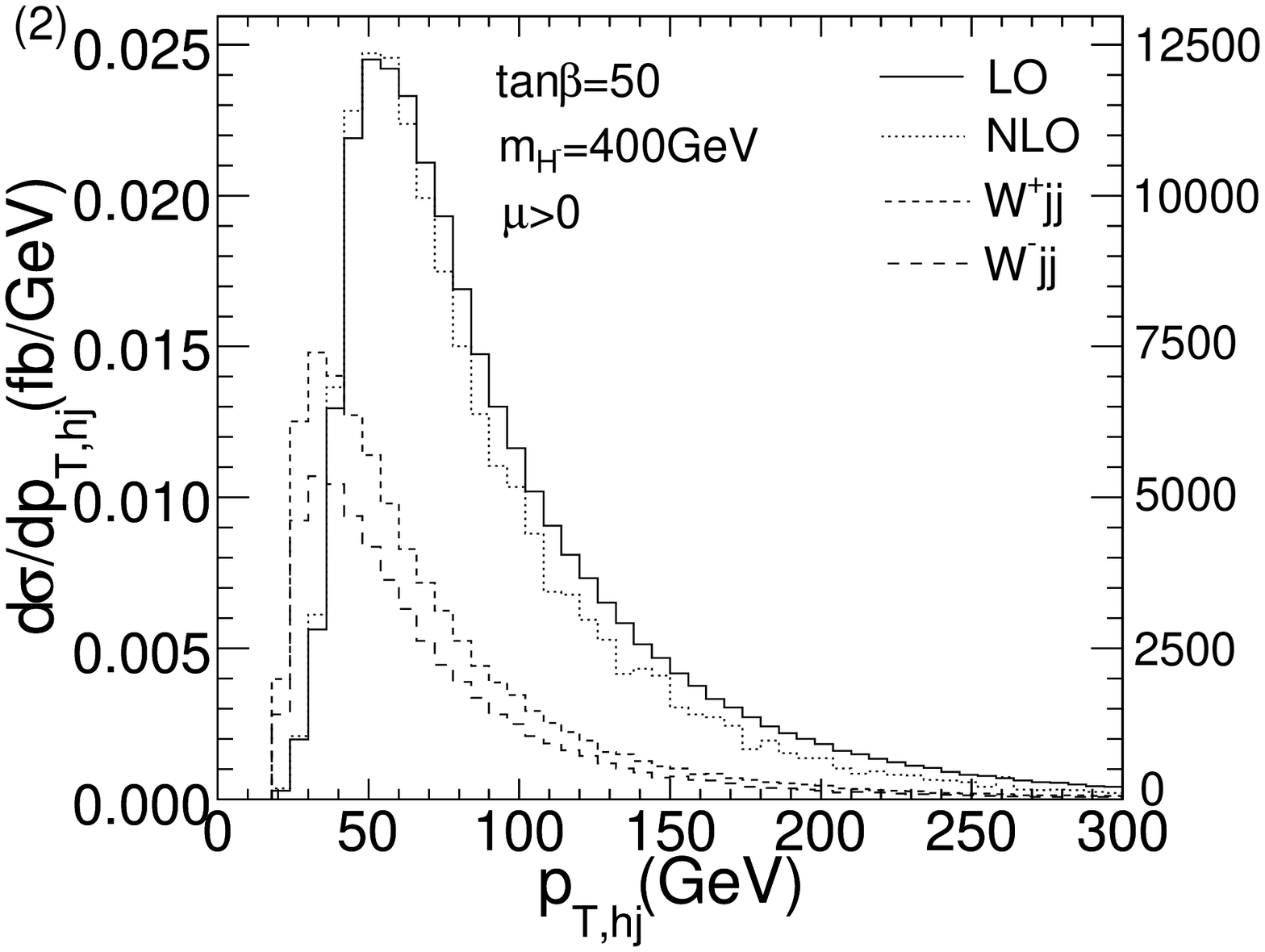}
\caption[]{Transverse momentum distributions of the hard light jet
for the signal and backgrounds after basic cuts, assuming: (1)
$m_0=200\ {\rm GeV},m_{1/2}=147\ {\rm GeV}$, and $A_{0}=200\ {\rm GeV}$;
(2) $m_0=320\ {\rm GeV},m_{1/2}=400\ {\rm GeV}$, and $A_{0}=300\ {\rm
GeV}$. The left axis scale is for the cross section
of the signal, while the right one is for the backgrounds.} \label{f14}
\end{figure}

\begin{figure}[!ht]
\includegraphics[width=0.80\textwidth]{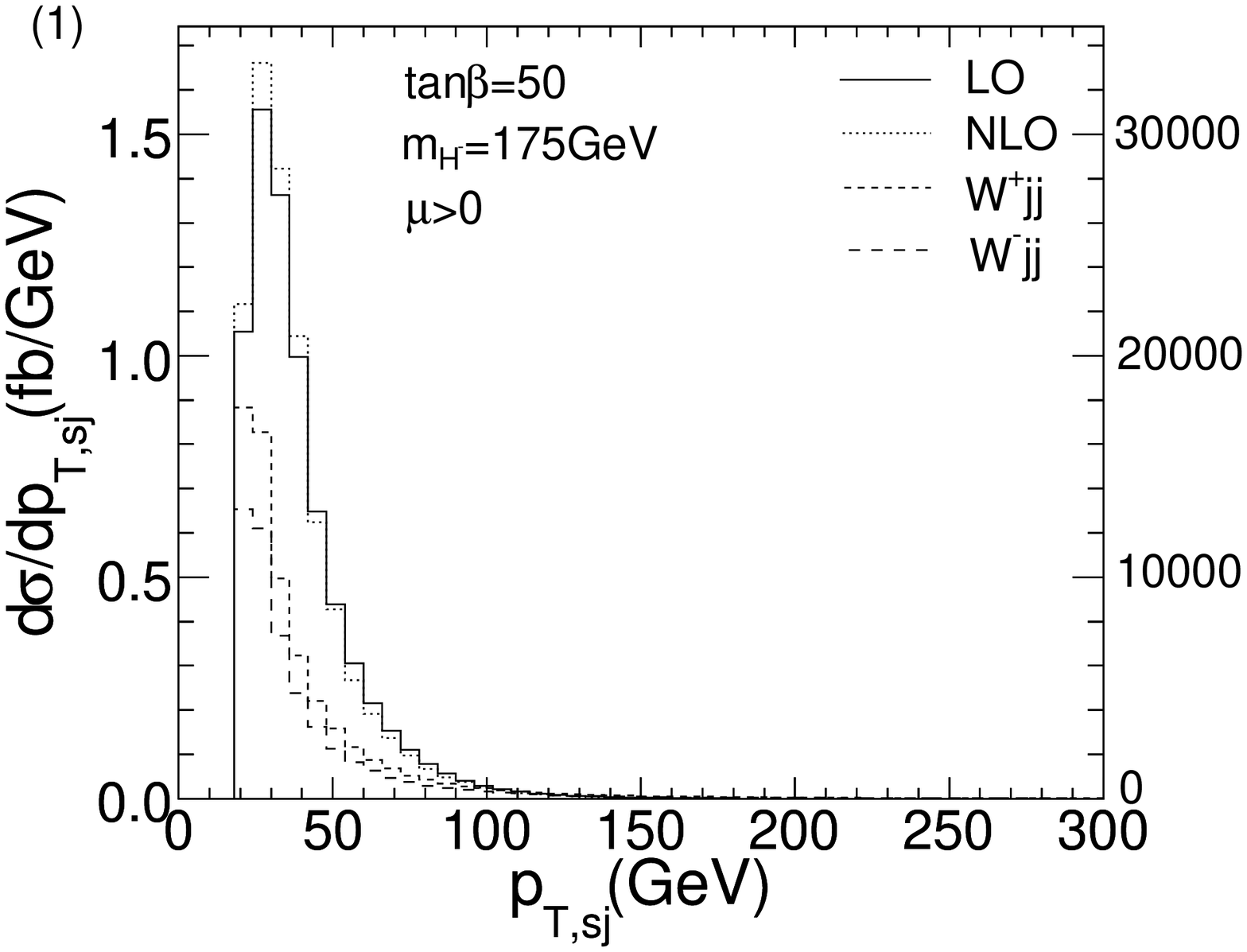}
\includegraphics[width=0.80\textwidth]{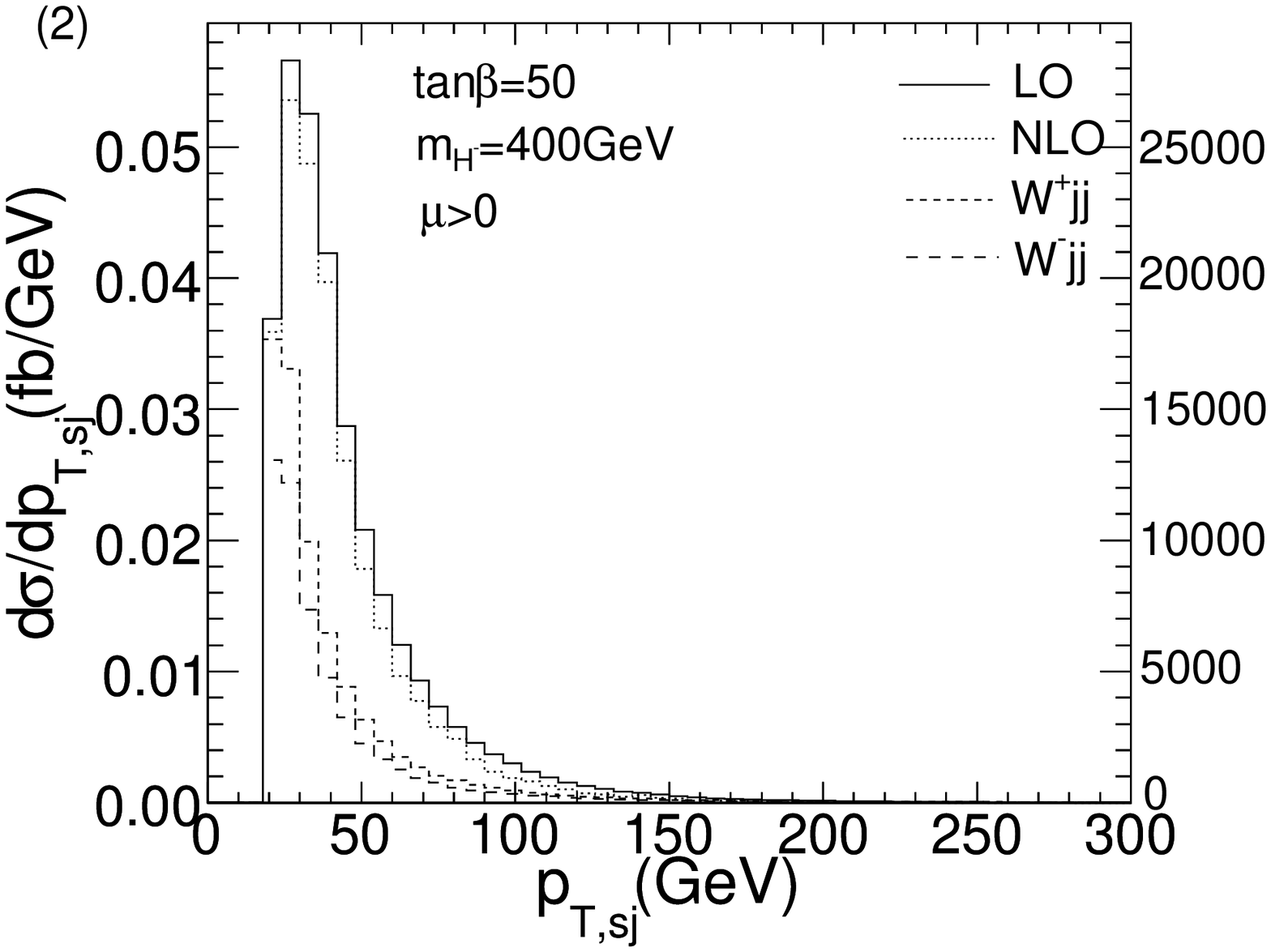}
\caption[]{Transverse momentum distributions of the soft light jet
for the signal and backgrounds after basic cuts, assuming: (1)
$m_0=200\ {\rm GeV},m_{1/2}=147\ {\rm GeV}$, and $A_{0}=200\ {\rm GeV}$;
(2) $m_0=320\ {\rm GeV},m_{1/2}=400\ {\rm GeV}$, and $A_{0}=300\ {\rm
GeV}$. The left axis scale is for the cross section
of the signal, while the right one is for the backgrounds.} \label{f15}
\end{figure}

Our simulation results for the relevant distributions are shown in
Figs.~\ref{f11}-\ref{f15}, which include the distributions of the
$m_T$, the $p_T$ for all jets and the missing transverse momentum
for the signal and backgrounds after the basic cuts, assuming: (1)
$m_0=\ 200{\rm GeV},m_{1/2}=147\ {\rm GeV},A_{0}=200\ {\rm GeV},\tan
\beta=50$ and $\mu>0$; (2) $m_0=320\ {\rm GeV},m_{1/2}=400\ {\rm
GeV},A_{0}=300\ {\rm GeV},\tan\beta=50$ and $\mu>0$. In case (1) of
those figures, the NLO QCD corrections can be neglected for all the
distributions, but in case (2), the NLO QCD corrections reduce the
LO results significantly, which can reach above 10\% in some region
of the distributions. It can be seen that the additional cuts
introduced at the LO still work well when including the NLO QCD
effects.

In the following calculations of the total cross sections the
additional cuts are used. Moreover, an integrated luminosity of $300
\ {\rm fb}^{-1}$ and a $\tau$ detection efficiency of $30\%$ are
taken to calculate the significance $S/ \sqrt{B}$. The total cross
sections for the backgrounds from the final state $W^+ +2j$ and $W^-
+2j$ are about 32 fb and 25 fb, respectively. Now, we add the cross
sections of the $H^-W^+$ and the $H^+W^-$ production together, as
well as for the backgrounds. Tables \ref{ta1} and \ref{ta2} show
some representative results of the cross sections and the
significance, where we can see that the significance can reach above
20 for $\tan \beta=50$ and $m_{H^-}=175\ \rm{GeV}$.
\begin{table}[!ht]
\begin{tabular}{|c|c|}
\hline
 &Integrated cross section (fb)\\
\qquad Parameter \qquad & \qquad \qquad Signal\qquad  Background \qquad $S/\sqrt{B}$\\
\hline
$m_{H^-}=175{\rm GeV}$ & LO \; \qquad 17.6\qquad \quad   \qquad 57 \qquad \quad \qquad 22.1 \;\\
& NLO \qquad 17.2\qquad \quad   \qquad 57 \qquad \quad \qquad 21.6 \;\\
\hline
$m_{H^-}=345{\rm GeV}$ & LO \; \qquad 2.12\qquad \quad   \qquad 57 \qquad \quad \qquad 2.66 \;\\
& NLO \qquad 1.84\qquad \quad   \qquad 57 \qquad \quad \qquad 2.31 \;\\
\hline
$m_{H^-}=630{\rm GeV}$ & LO \; \qquad 0.34\qquad \quad   \qquad 57 \qquad \quad \qquad 0.43 \;\\
& NLO \qquad 0.28\qquad \quad   \qquad 57 \qquad \quad \qquad 0.35 \;\\
\hline
\end{tabular}\caption{Results under several groups of $m_{H^-}$ value,
assuming $m_0=A_0=200 \ {\rm GeV}$, and $\tan\beta=50$.} \label{ta1}
\end{table}
\begin{table}[!ht]
\begin{tabular}{|c|c|}
\hline
 &Integrated cross section (fb)\\
\qquad Parameter \qquad & \qquad \qquad Signal\qquad  Background \qquad $S/\sqrt{B}$\\
\hline
$\tan\beta=20$ & LO \; \qquad 0.46\qquad \quad   \qquad 57 \qquad \quad \qquad 0.58 \;\\
& NLO \qquad 0.42\qquad \quad   \qquad 57 \qquad \quad \qquad 0.53 \;\\
\hline
$\tan\beta=40$ & LO \; \qquad 4.60\qquad \quad   \qquad 57 \qquad \quad \qquad 5.78 \;\\
& NLO \qquad 4.34\qquad \quad   \qquad 57 \qquad \quad \qquad 5.45 \;\\
\hline
$\tan\beta=50$ & LO \; \qquad 17.7\qquad \quad   \qquad 57 \qquad \quad \qquad 22.3 \;\\
& NLO \qquad 16.7\qquad \quad   \qquad 57 \qquad \quad \qquad 21.0 \;\\
\hline
\end{tabular}\caption{Results under several groups of $\tan\beta$ value,
assuming $m_0=A_0=200 \ {\rm GeV}$, and $m_{1/2}=150 \ {\rm GeV}$.
The three $\tan\beta$ values correspond to $m_{H^{\pm}}=282\ {\rm GeV}$, $224\ {\rm GeV}$
and $178\ {\rm GeV}$ from top to bottom, respectively.}
\label{ta2}
\end{table}

\begin{figure}[!ht]
\includegraphics[width=0.90\textwidth]{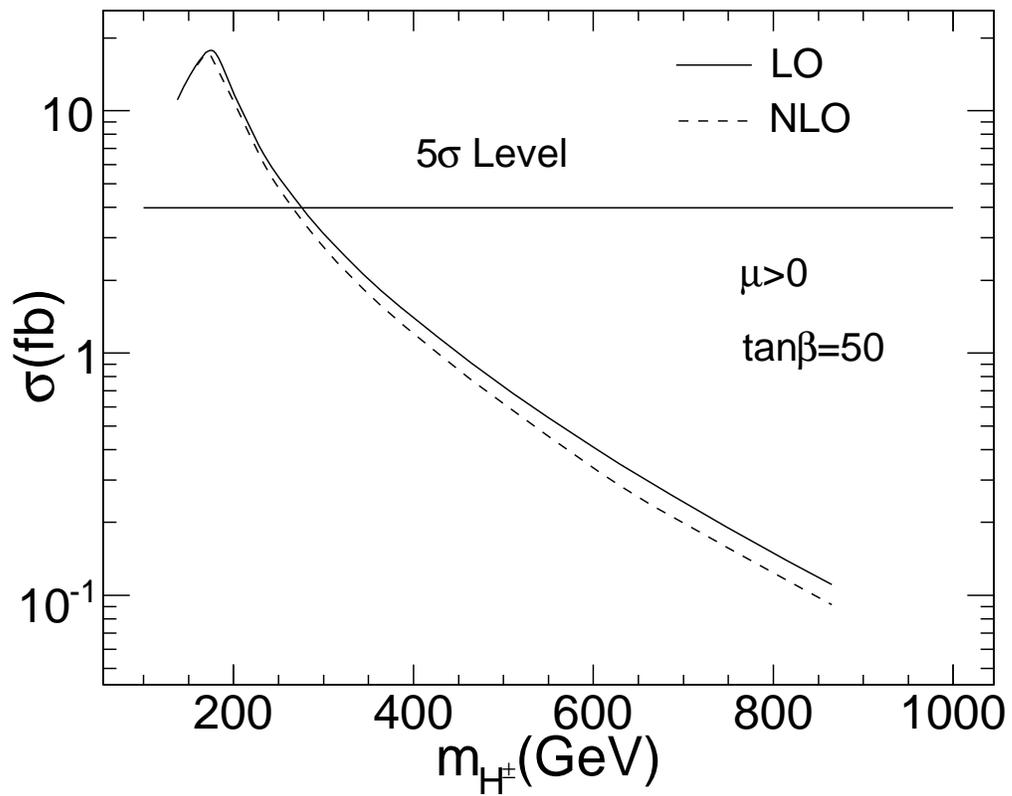}
\caption[]{$H^{\pm}$ mass dependence of the integrated cross section
after all cuts, assuming $m_0=A_0=200\ {\rm GeV}$.} \label{f16}
\end{figure}

\begin{figure}[!ht]
\includegraphics[width=0.90\textwidth]{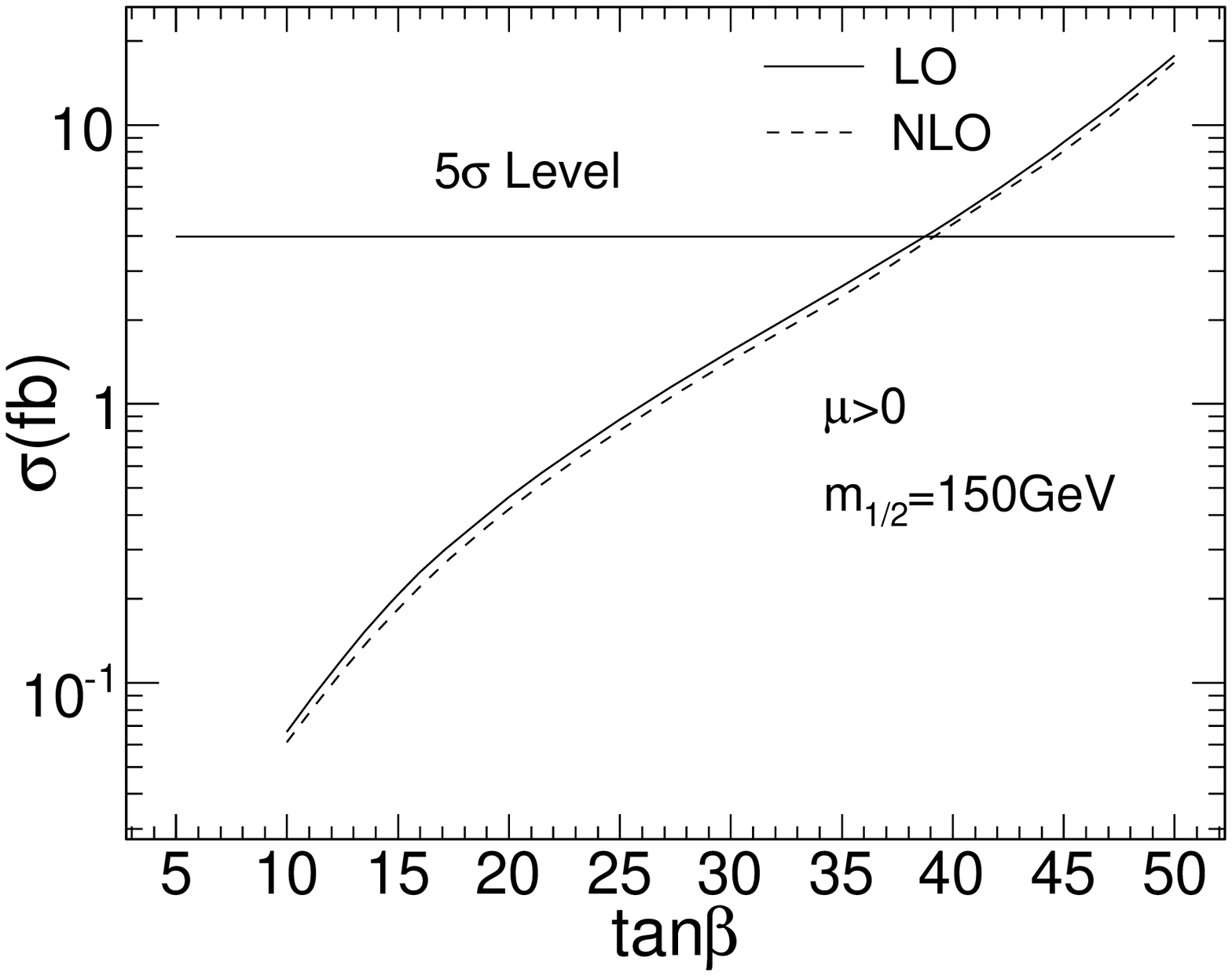}
\caption[]{$\tan \beta$ dependence of the integrated cross section
after all cuts, assuming $m_0=A_0=200\ {\rm GeV}$.} \label{f17}
\end{figure}

Figs.~\ref{f16} and \ref{f17} show the dependence of the cross
sections on $m_{H^-}$ mass and $\tan\beta$, respectively. In
general, the NLO QCD corrections reduce the cross sections in most
of the parameter space, and their magnitude can be larger than 10\%.
The horizontal lines in the figures correspond to the total cross
sections required for $S/\sqrt{B}=5$. It can be seen that at the LO
the signal can be detected at a $5\sigma$ level for $\tan\beta
\gtrsim 40$ and $m_{1/2}=150 \ {\rm GeV}$, and for $100 \ {\rm GeV}
\lesssim m_{H^-} \lesssim 250 \ {\rm GeV}$ and $\tan \beta=50$,
respectively. And the NLO QCD corrections have small effects on the
above results. Some of our results are different from those given in
Ref.~\cite{Eriksson:2006yt} mainly due to the difference between the
mSUGRA scenario and the one used in Ref.~\cite{Eriksson:2006yt}.
\section{CONCLUSIONS}
In conclusion, we have calculated the NLO QCD corrections to the
inclusive total cross sections of the $H^{\mp}W^{\pm}$ pairs
produced at the LHC through $b\bar {b}$ annihilation in the MSSM.
The NLO QCD corrections can either enhance or reduce the total cross
sections, but they generally efficiently reduce the dependence of
the total cross sections on the renormalization/factorization scale.
The magnitude of the NLO QCD corrections is about 10\% in most of
the parameter space and can reach 15\% in some parameter region.
Finally, we give some discussion on the $H^{\mp}W^{\pm} \to
2j+\tau_{jet}+\not{\!p}_T$ signal including the NLO QCD effects, and
find an observable signal at a $5\sigma$ level in some region of the
mSUGRA parameter space.
\begin{acknowledgments}
This work was supported in part by the National Natural Science Foundation of China,
under Grants No.10421503, No.10575001 and No.10635030 and the Key Grant Project of Chinese
Ministry of Education under Grant No.305001.
\end{acknowledgments}

\section*{APPENDIX}
\begin{figure}[!ht]
\includegraphics[width=0.90\textwidth]{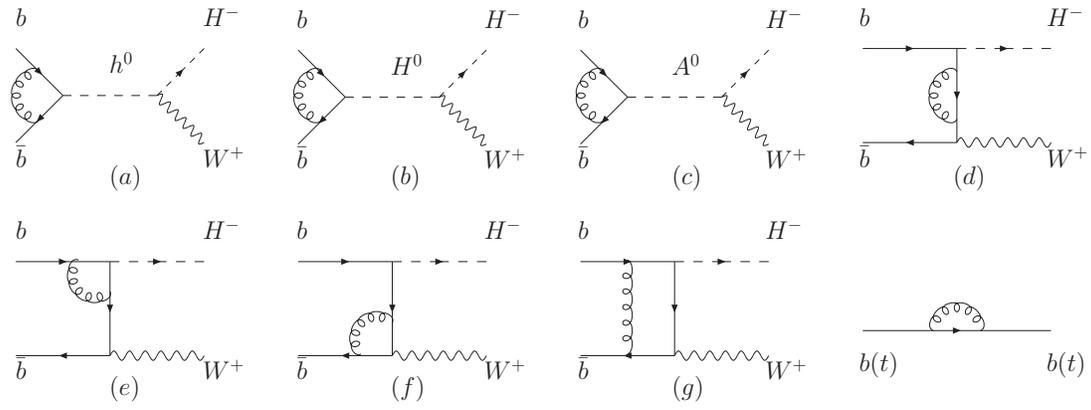}
\caption[]{One-loop virtual Feynman diagrams for $b\bar{b} \to H^-W^+$.}
\label{f18}
\end{figure}

In this appendix, we give the relevant Feynman rules and the form
factors for the virtual amplitude.
First we give the relevant Feynman rules.\\
1. $h^0(H^0)-b-\bar{b}:\quad {\mathcal A}_{1(2)}m_b$
\begin{equation}
{\mathcal A}_1=\frac{ie\delta_{ij}s_{\alpha}}{2c_{\beta}M_Ws_W},\quad
{\mathcal A}_2=\frac{-ie\delta_{ij}c_{\alpha}}{2c_{\beta}M_Ws_W},\nonumber
\end{equation}
where $\alpha$ is the mixing angle in the CP even neutral Higgs boson sector.
Here we use the abbreviations $s_{\alpha}=\sin \alpha$, $s_{\beta}=\sin \beta$
and so on.\\
2. $A^0-b-\bar{b}:\quad {\mathcal A}_3m_b\gamma_5$
\begin{equation}
{\mathcal A}_3=\frac{-e\delta_{ij}t_{\beta}}{2M_Ws_W}.\nonumber
\end{equation}\\
3. $h^0(H^0,A^0)-W^--H^+:\quad {\mathcal F}_{1(2,3)}(p_{h^0(H^0,A^0)}-p_{H^+})^{\mu}$
\begin{equation}
{\mathcal F}_1=\frac{iec_{\beta-\alpha}}{2s_W},\quad {\mathcal F}_2=-\frac{ies_{\beta-\alpha}}{2s_W},
\quad {\mathcal F}_3=\frac{e}{2s_W}.\nonumber
\end{equation}
Here we define the ingoing four-momenta to be positive.\\
4. $b-\bar{t}-H^+:\quad {\mathcal G}_1P_L+{\mathcal G}_2P_R$
\begin{equation}
{\mathcal G}_1=\frac{ie\delta_{ij}m_t}{\sqrt{2}M_Ws_Wt_{\beta}},\quad
{\mathcal G}_2=\frac{ie
\delta_{ij}m_bt_{\beta}}{\sqrt{2}M_Ws_W}.\nonumber
\end{equation}
Here and below, we assume the third generation CKM matrix
element $V_{tb}$ equal to 1.\\
5. $\bar{b}-t-W^-:\quad {\mathcal G}_3\gamma^{\mu}P_L$
\begin{equation}
{\mathcal G}_3=\frac{-ie}{\sqrt{2}s_W}.\nonumber
\end{equation}\\
Below we collect the explicit expressions of the nonzero form
factors in Eq.(\ref{e5}). For simplicity, we introduce the following
abbreviations for the Passarino-Veltman two-point integrals
$B_{i(j)}$, three-point integrals $C_{i(j)}$ and four-point
integrals $D_{i(j)}$, which are defined similar to
Ref.~\cite{Denner:1991kt}
except that we take internal masses squared as arguments:\\
\noindent $B^a_{i(j)}=B_{i(j)}(s,0,0)$,
\\
$B^b_{i(j)}=B_{i(j)}(t,0,m_t^2)$,
\\
$C^a_{i,(j)}=C_{i(j)}(M_{H^-}^2,0,t,m_t^2,0,0)$,
\\
$C^b_{i,(j)}=C_{i(j)}(M_W^2,0,t,m_t^2,0,0)$,
\\
$C^c_{i,(j)}=C_{i(j)}(0,s,0,0,0,0)$,
\\
$C^d_{i,(j)}=C_{i(j)}(0,t,M_{H^-}^2,0,0,m_t^2)$,
\\
$C^e_{i,(j)}=C_{i(j)}(0,t,M_W^2,0,0,m_t^2)$,
\\
$D_{i(j)}=D_{i(j)}(M_W^2,s,0,t,M_{H^-}^2,0,m_t^2,0,0,0)$.
\\
Most of the above functions contain IR singularities. Since all the
Passarino-Veltman integrals can be written as a combination of the scalar
functions $A_0,B_0,C_0$ and $D_0$, we present here the explicit expressions for the
$C_0$ and $D_0$ functions, which contain the IR divergences and were used in our calculations:\\
$\displaystyle{C^c_0=\frac{C_{\epsilon}}{s}\left[\frac{1}{\epsilon^2}-\frac{\pi^2}{3}\right],}$\\
$\displaystyle{C^a_0=C^d_0=\frac{C_{\epsilon}}{t-M_{H^-}^2}\left[\frac{1}{\epsilon}\ln\frac{m_t^2-M_{H^-}^2}
{m_t^2-t}+{\rm Li}\left(\frac{M_{H^-}^2}{M_{H^-}^2-m_t^2}\right)
-{\rm Li}\left(\frac{t}{t-m_t^2}\right) \right.}$\\
$\displaystyle{\qquad \left.-\frac{1}{2}\ln^2(m_t^2-M_{H^-}^2)+\frac{1}{2}\ln^2(m_t^2-t)+\ln s \ln \frac
{m_t^2-M_{H^-}^2}{m_t^2-t}\right],}$\\
$\displaystyle{C^e_0=C^b_0=C^a_0(M_W^2 \leftrightarrow M_{H^-}^2),}$\\
$\displaystyle{D_0=\frac{C_{\epsilon}}{s(t-m_t^2)}\left[\frac{1}{\epsilon ^2}-\frac{1}{\epsilon}
\ln \frac{(t-m_t^2)^2}{(m_t^2-M_{H^-}^2)(m_t^2-M_W^2)}\right.}$\\
$\displaystyle{\qquad -\ln s \ln\frac{(t-m_t^2)^2}{(m_t^2-M_{H^-}^2)(m_t^2-M_W^2)}
-2{\rm Li}\left(1+\frac{m_t^2-M_W^2}{t-m_t^2}\right)
-2{\rm Li}\left(1+\frac{m_t^2-M_{H^-}^2}{t-m_t^2}\right)}$\\
$\displaystyle{\qquad -{\rm Li}\left[1+\frac{(m_t^2-M_{H^-}^2)(m_t^2-M_W^2)}{sm_t^2}\right]
-\frac{1}{2}\ln ^2\left(\frac{s}{m_t^2}\right)+2\ln s\ln\left(\frac{t-m_t^2}{m_t^2}
\right)}$\\
$\displaystyle{\qquad \left.-\ln(m_t^2-M_{H^-}^2)\ln\left(\frac{m_t^2-M_{H^-}^2}{m_t^2}\right)
-\ln(m_t^2-M_W^2)\ln\left(\frac{m_t^2-M_W^2}{m_t^2}\right)\right],}$\\
where $C_{\epsilon}=(4\pi\mu_r^2/s)^{\epsilon}\Gamma(1-\epsilon)/\Gamma(1-2\epsilon)$.
For diagrams(a)-(f) in Fig.\ref{f18}, we get the form factors as following, respectively,\\
$\displaystyle{f_5^a=f_6^a=f_9^a=f_{10}^a=C_F\delta_{ij}\frac{\alpha_s}{4\pi}
\frac{e^2m_bs_{\alpha}c_{\beta-\alpha}}{c_{\beta}M_{W}s_W^2(s-m_{h^0}^2)}
\left[(C_0^c+C_1^c+C_2^c)s-2B_0^a\right],}$\\
$\displaystyle{f_5^b=f_6^b=f_9^b=f_{10}^b=C_F\delta_{ij}\frac{\alpha_s}{4\pi}
\frac{e^2m_bc_{\alpha}s_{\beta-\alpha}}{c_{\beta}M_{W}s_W^2(s-m_{H^0}^2)}
\left[(C_0^c+C_1^c+C_2^c)s-2B_0^a\right],}$\\
$\displaystyle{f_5^c=f_9^c=-f_6^c=-f_{10}^c=C_F\delta_{ij}\frac{\alpha_s}{4\pi}
\frac{e^2m_bt_{\beta}}{M_{W}s_W^2(s-m_{A^0}^2)}
\left[(C_0^c+C_1^c+C_2^c)s-2B_0^a\right],}$\\
$\displaystyle{f_2^d=-C_F\delta_{ij}\frac{\alpha_s}{2\pi}\frac{e}{\sqrt{2}s_W(t-m_t^2)^2}
(2m_ta)(tB^b_1-m_t^2B^b_0),}$\\
$\displaystyle{f_9^d=2f_3^d=-C_F\delta_{ij}\frac{\alpha_s}{2\pi}\frac{e}{\sqrt{2}s_W(t-m_t^2)^2}
(-2b)[m_t^2(B^b_1-3B^b_0)+t(B^b_0+B^b_1)],}$\\
$\displaystyle{f_2^e=C_F\delta_{ij}\frac{\alpha_s}{4\pi}\frac{e}{\sqrt{2}s_W(t-m_t^2)}
(am_t)\left[4B_0^b+2(M_{H^-}^2-t)C_1^d+2(2M_{H^-}^2-t)C_2^d\right],}$\\
$\displaystyle{f_9^e=2f_3^e=C_F\delta_{ij}\frac{\alpha_s}{4\pi}\frac{e}{\sqrt{2}s_W(t-m_t^2)}
(-2b)\left[4B_0^b+2(M_{H^-}^2-t)C_1^d+2(2M_{H^-}^2-m_t^2)C_2^d\right],}$\\
$\displaystyle{f_2^f=C_F\delta_{ij}\frac{\alpha_s}{4\pi}\frac{e}{\sqrt{2}s_W(t-m_t^2)}
(2am_t)\left[2C_{00}^e+M_W^2(C_2^e+C_{22}^e)+(M_W^2-t)(C_1^e+C_{12}^e)\right],}$\\
$\displaystyle{f_3^f=C_F\delta_{ij}\frac{\alpha_s}{4\pi}\frac{e}{\sqrt{2}s_W(t-m_t^2)}
(-2b)\left[2C_{00}^e+M_W^2(C_2^e+C_{22}^e)+(M_W^2-t)(C_1^e+C_{12}^e)\right],}$\\
$\displaystyle{f_9^f=C_F\delta_{ij}\frac{\alpha_s}{4\pi}\frac{e}{\sqrt{2}s_W(t-m_t^2)}
(-4b)\left[2C_{00}^e+M_W^2(C_1^e+C_2^e+C_{12}^e+C_{22}^e)-m_t^2C_1^e\right],}$\\
$\displaystyle{f_{12}^f=C_F\delta_{ij}\frac{\alpha_s}{4\pi}\frac{e}{\sqrt{2}s_W(t-m_t^2)}
C_{12}^e,}$\\
where a,b are abbreviations for
\begin{equation}
a=\frac{em_t}{\sqrt{2}M_Ws_Wt_{\beta}},\qquad b=\frac{em_bt_{\beta}}{\sqrt{2}M_Ws_W}.\nonumber
\end{equation}
For the box diagram(g) in Fig.\ref{f18}, we find\\
$\displaystyle{f_2^{box}=C_F\delta_{ij}\frac{\alpha_s}{4\pi}\frac{e}{\sqrt{2}s_W}(-2m_ta)\left\{\left[
-2D_{00}-M_W^2(D_{11}+D_{13}+D_1+D_3) \right. \right. }$\\
$\displaystyle{\left. \left. \qquad -M_{H^-}^2(D_2+D_3+D_{22}+D_{23})\right]
-(D_{12}+D_{13}+D_{23}+D_{33})t-(D_{12}+D_0+D_1+D_2)u\right\},}$\\
$\displaystyle{f_3^{box}=C_F\delta_{ij}\frac{\alpha_s}{4\pi}\frac{e}{\sqrt{2}s_W}(-2b)D_3s,}$\\
$\displaystyle{f_5^{box}=C_F\delta_{ij}\frac{\alpha_s}{4\pi}\frac{e}{\sqrt{2}s_W}(4b)
\left[-C^a_0+M^2_W(D_0+D_1+D_2+D_3)-D_2s\right],}$\\
$\displaystyle{f_8^{box}=C_F\delta_{ij}\frac{\alpha_s}{4\pi}\frac{e}{\sqrt{2}s_W}(-2m_ta)
\left[-2(D_0+D_1+2D_2+D_3+D_{12}+D_{22}+D_{23})\right],}$\\
$\displaystyle{f_9^{box}=C_F\delta_{ij}\frac{\alpha_s}{4\pi}\frac{e}{\sqrt{2}s_W}(4b)
\left[C_2^b-C_0^a+M_W^2(D_0+D_1)+(t+u-M^2_{H^-})D_2+(t+u)D_3\right],}$\\
$\displaystyle{f_{12}^{box}=C_F\delta_{ij}\frac{\alpha_s}{4\pi}\frac{e}{\sqrt{2}s_W}(-2m_ta)
\left[-2(D_2+D_3+D_{12}+D_{13}+D_{22}+2D_{23}+D_{33})\right].}$\\
\newpage
\bibliography{gaugino}
\newpage

\end{document}